\documentclass[iop]{emulateapj}

\shorttitle{On the universal late X-ray emission of BdHNe and its possible collimation}
\shortauthors{Pisani et al.}

\begin{document}

\title{On the universal late X-ray emission of binary-driven hypernovae and its possible collimation}

\author{G.~B. Pisani\altaffilmark{1,2}, R. Ruffini\altaffilmark{1,2,3,4}, Y. Aimuratov\altaffilmark{1,3}, C.~L. Bianco\altaffilmark{1,2}, M. Kovacevic\altaffilmark{1,3},  R. Moradi\altaffilmark{1,2}, M. Muccino\altaffilmark{1,2}, A.~V. Penacchioni\altaffilmark{5,6}, J.~A. Rueda\altaffilmark{1,2,4}, S. Shakeri\altaffilmark{2,7}, Y.~Wang\altaffilmark{1,2}}

\altaffiltext{1}{Dipartimento di Fisica, Sapienza Universit\`a  di Roma and ICRA, Piazzale Aldo Moro 5, I-00185 Roma, Italy}
\altaffiltext{2}{ICRANet, Piazza della Repubblica 10, I-65122 Pescara, Italy}
\altaffiltext{3}{Universit\'e de Nice Sophia-Antipolis, Grand Ch\^ateau Parc Valrose, Nice, CEDEX 2, France}
\altaffiltext{4}{ICRANet-Rio, Centro Brasileiro de Pesquisas Fisicas, Rua Dr. Xavier Sigaud 150, Rio de Janeiro, RJ, 22290-180, Brazil}
\altaffiltext{5}{University of Siena, Department of Physical Sciences, Earth and Environment, Via Roma 56, I-53100 Siena, Italy}
\altaffiltext{6}{ASI Science Data Center, via del Politecnico s.n.c., I-00133 Rome Italy}
\altaffiltext{7}{Department of Physics, Isfahan University of Technology, Isfahan 84156-83111, Iran}

\begin{abstract}
It has previously been discovered that there is a universal power law behavior exhibited by the late X-ray emission (LXRE) of a ``golden sample'' (GS) of six long energetic GRBs, when observed in the rest-frame of the source.
This remarkable feature, independent of the different isotropic energy ($E_{iso}$) of each GRB, has been used to estimate the cosmological redshift of some long GRBs.
This analysis is extended here to a new class of 161 long GRBs, all with $E_{iso}>10^{52}$ erg.
These GRBs are indicated as binary-driven hypernovae (BdHNe) in view of their progenitors: a tight binary system composed of a carbon-oxygen core (CO$_{\mathrm{core}}$) and a neutron star undergoing an induced gravitational collapse (IGC) to a black hole triggered by the CO$_{\mathrm{core}}$ explosion as a supernova (SN).
We confirm the universal behavior of the LXRE for the ``enlarged sample'' (ES) of 161 BdHNe observed up to the end of 2015, assuming a double-cone emitting region. We obtain a distribution of half-opening angles peaking at $\theta=17.62^\circ$, with a mean value of $30.05^\circ$, and a standard deviation of $19.65^\circ$.
This, in turn, leads to the possible establishment of a new cosmological candle.
Within the IGC model, such universal LXRE behavior is only indirectly related to the GRB and originates from the SN ejecta, of a standard constant mass, being shocked by the GRB emission.
The fulfillment of the universal relation in the LXRE and its independence of the prompt emission, further confirmed in this article, establishes a crucial test for any viable GRB model.
\end{abstract}


\section{Introduction}

The initial observations by the BATSE instrument on board the Compton $\gamma$-ray Observatory satellite have evidenced what has later become known as the prompt radiation of GRBs. On the basis of their hardness as well as their duration, GRBs were initially classified into short and long at a time when their cosmological nature was still being disputed \citep{1981Ap&SS..80....3M,Klebesadel1992,Dezalay1992,Koveliotou1993,Tavani1998}.

\begin{figure}
\centering
\includegraphics[width=\hsize,clip]{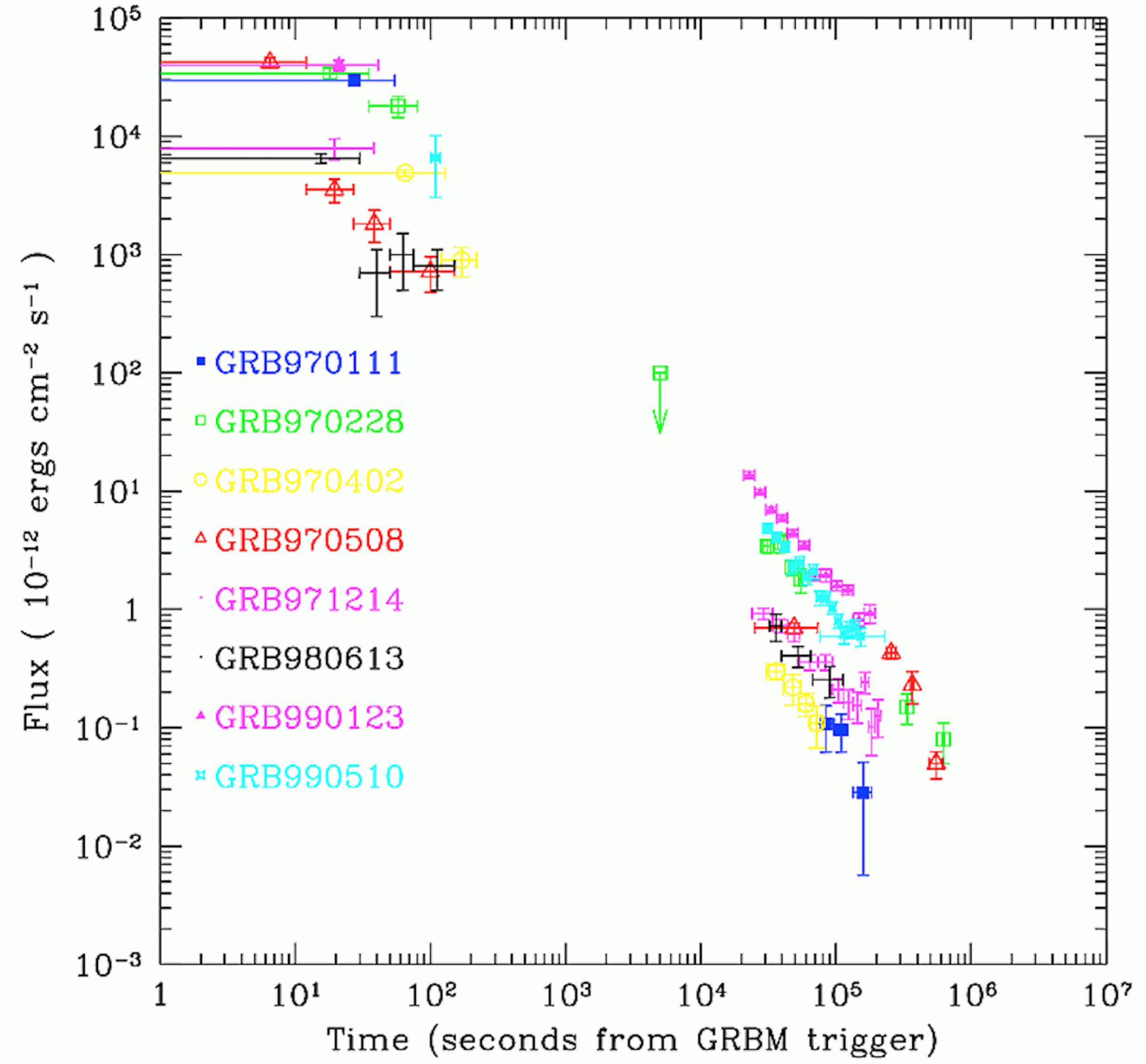}
\caption{Collection of X-ray afterglow light curves observed by the Italian-Dutch satellite \textit{BeppoSAX}.}
\label{BeppoSAX}
\end{figure}

The advent of the \textit{BeppoSAX} satellite \citep{1997AeAS..122..299B} introduced a novel approach to GRBs by introducing joint observations in the X-rays and $\gamma$-rays thanks to its instruments: the Gamma-ray Burst Monitor ($40$--$700$ keV), the Wide Field Cameras ($2$--$26$ keV), and the Narrow Field Instruments ($2$-$10$ keV). The unexpected and welcome discovery of the existence of a well separate component in the GRB soon appeared: the afterglow radiation lasting up to $10^5$--$10^6$ s after the emission of the prompt radiation \citep[see][]{1997Natur.387..783C,1997IAUC.6576....1C,1998ApJ...493L..67F,2000ApJS..127...59F,2006A&A...455..813D}. Beppo-SAX clearly indicated the existence of a power law behavior in the late X-ray emission (LXRE; see Fig. \ref{BeppoSAX}).

The coming of the \textit{Swift} satellite \citep{2004ApJ...611.1005G, 2007A&A...469..379E, 2010A&A...519A.102E}, significantly extending the observation in the X-ray band thanks to its X-ray Telescope (XRT band: $0.3$--$10$ keV), has allowed us for the first time to cover the unexplored region between the end of the prompt radiation and the power law late X-ray behavior discovered by \textit{BeppoSAX}: in some long GRBs a steep decay phase was observed followed by a plateau leading then to a typical LXRE power law behavior \citep{2007A&A...469..379E, 2010A&A...519A.102E}.

\begin{figure}
\centering
\includegraphics[width=\hsize,clip]{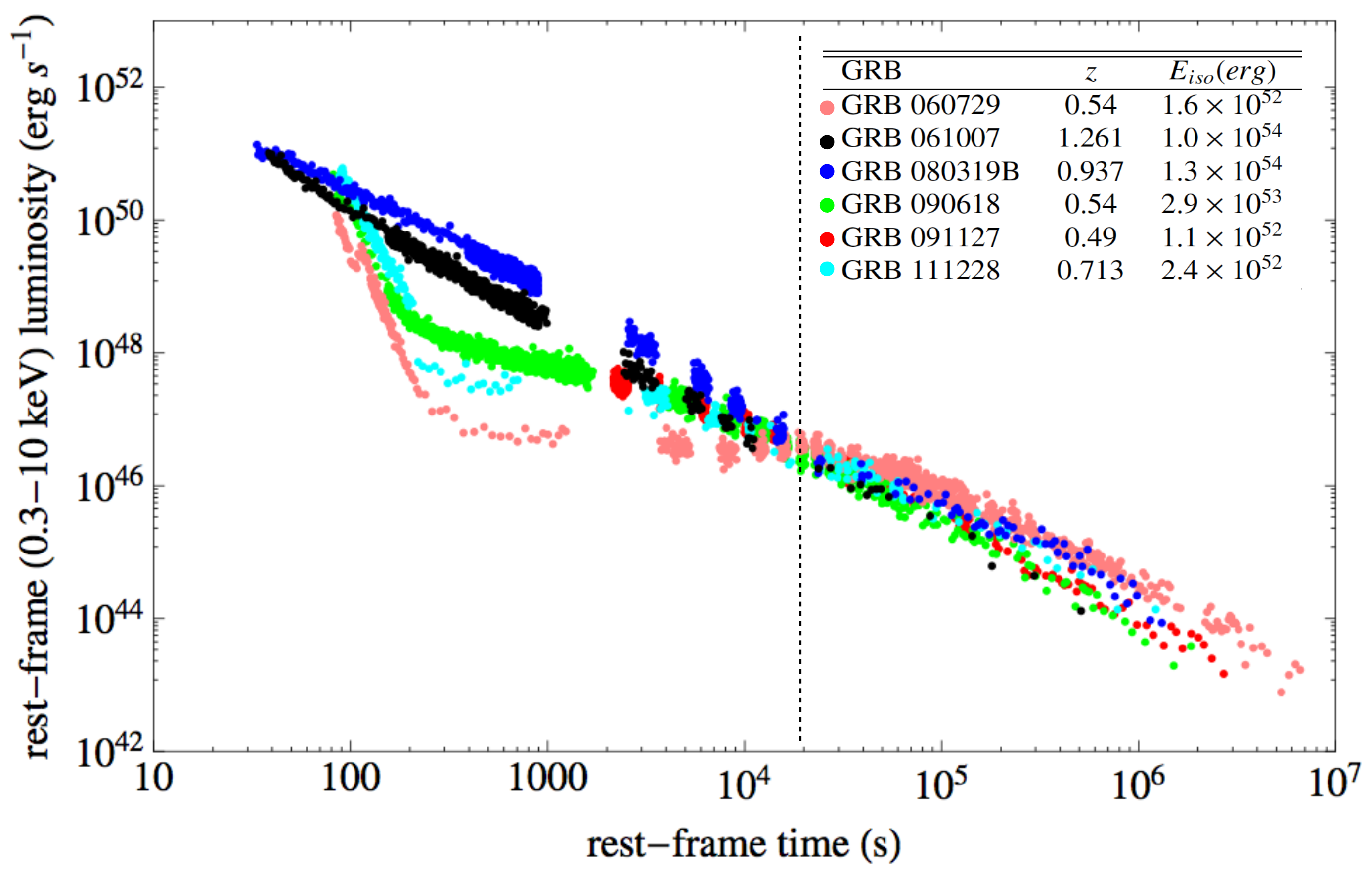}
\caption{Scaling law found in the isotropic X-ray late times luminosity within the GS by \citet{Pisani2013}. Despite the different early behavior, the different light curves join all together the same power law after a rest-frame time of $t_{rf}\sim2\times10^4$ s.}
\label{scaling}
\end{figure}

\begin{figure}
\centering
\includegraphics[width=\hsize,clip]{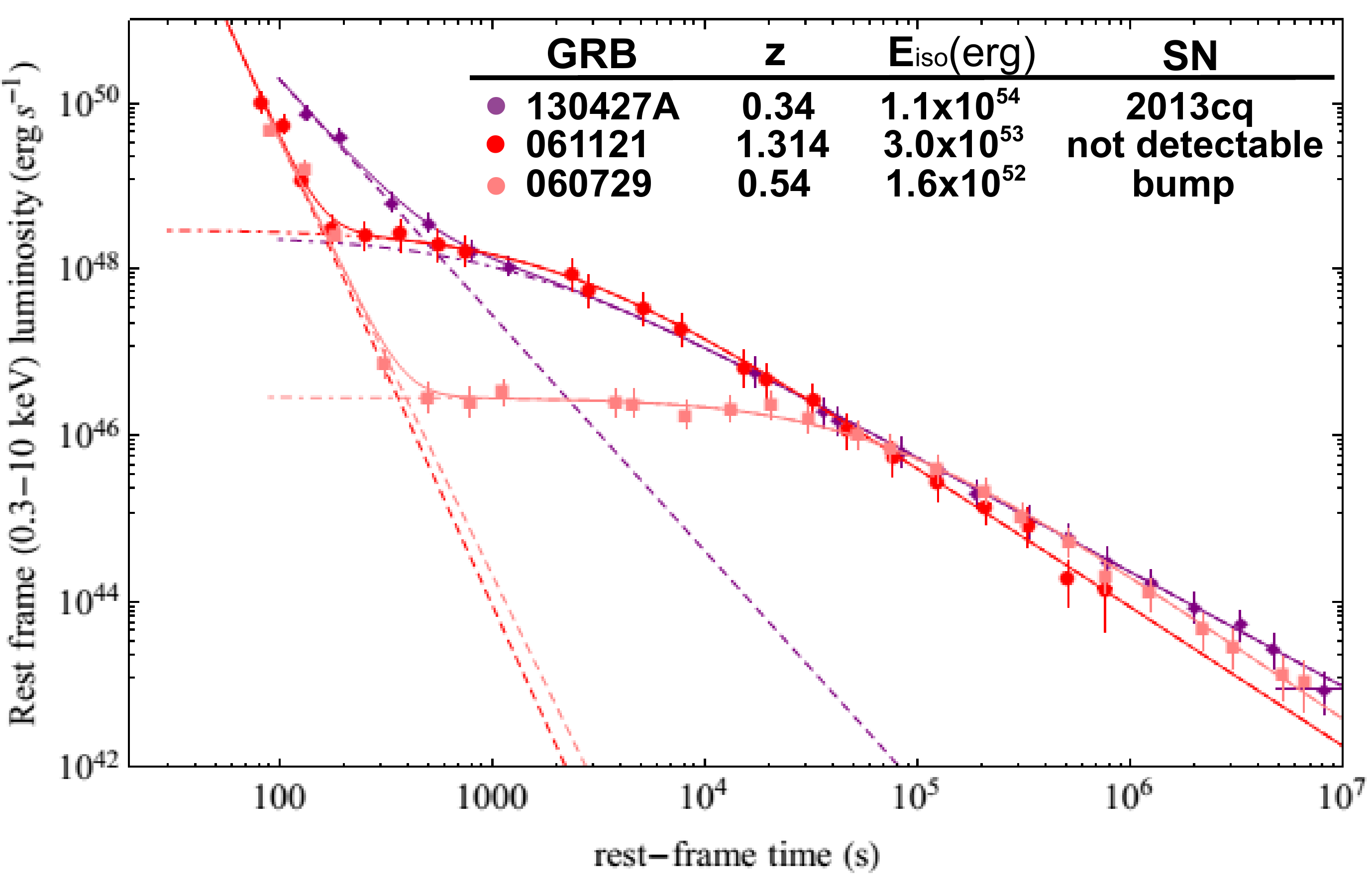}
\caption{Nested structure of the isotropic X-ray luminosity of the BdHNe. This includes the previously mentioned scaling law of the late power law and leads to an inverse proportionality between the luminosity of the plateau and the rest-frame time delimiting its end and the beginning of the late power law decay \citet{2014A&A...565L..10R}.}
\label{nesting}
\end{figure}

Already, \citet{Pisani2013} noticed the unexpected result that the LXREs of a ``golden sample'' (GS) of six long, closeby ($z \lesssim 1$), energetic ($E_{iso} > 10^{52}$ erg) GRBs, when measured in the rest-frame of the sources, were showing a common power law behavior (see Fig. \ref{scaling}), independently from the isotropic energy $E_{iso}$ coming from the prompt radiation (see Fig \ref{nesting}).
More surprising was the fact that the plateau phase luminosity and duration before merging in the common LXRE power law behavior were clearly functions of the $E_{iso}$ \citep[see Fig. \ref{nesting}, and][]{2014A&A...565L..10R}, while the late power law remains independent from the energetic of the prompt radiation \citep[see Fig. \ref{scaling}--\ref{nesting}, and][]{Pisani2013,2014A&A...565L..10R}.
For this reason, this remarkable scaling law has been used as a standard candle to independently estimate the cosmological redshift of some long GRBs by requiring the overlap of their LXRE \citep[see, e.g.,][]{Penacchioni2012,Ana2013,2013GCN..14888...1R,2013GCN..15576...1R,2014GCN..15707...1R}, and also to predict, 10 days in advance, the emergence of the typical optical signature of the supernova SN 2013cq, associated with GRB 130427A \citep{2015ApJ...798...10R,2013GCN..14526...1R,2013GCN..14646...1D,2013GCN..14686...1L}.

The current analysis is based on the paradigms introduced in \citet{2001ApJ...555L.107R} for the spacetime parametrization of the GRBs, in \citet{2001ApJ...555L.113R} for the interpretation of the structure of the GRB prompt emission, and in \citet{2001ApJ...555L.117R} for the induced gravitational collapse (IGC) process, further evolved in \citet{Ruffini2007b},\citet{Rueda2012}, \citet{2014ApJ...793L..36F}, and \citet{2016arXiv160202732R}.
In the present case, the phenomenon points to an IGC occurring when a tight binary system composed of a carbon-oxygen core (CO$_{\mathrm{core}}$) undergoes a supernova (SN) explosion in the presence of a binary neutron star (NS) companion \citep{2001ApJ...555L.113R, Ruffini2007b, Izzo2012b, Rueda2012, 2014ApJ...793L..36F, 2015ApJ...798...10R}.
When the IGC leads the NS to accrete enough matter and therefore to collapse to a black hole (BH), the overall observed phenomenon is called binary-driven hypernova \citep[BdHN; ][]{2014ApJ...793L..36F, 2015ApJ...798...10R, 2016arXiv160202732R}.

\begin{figure}
\centering
\includegraphics[width=\hsize,clip]{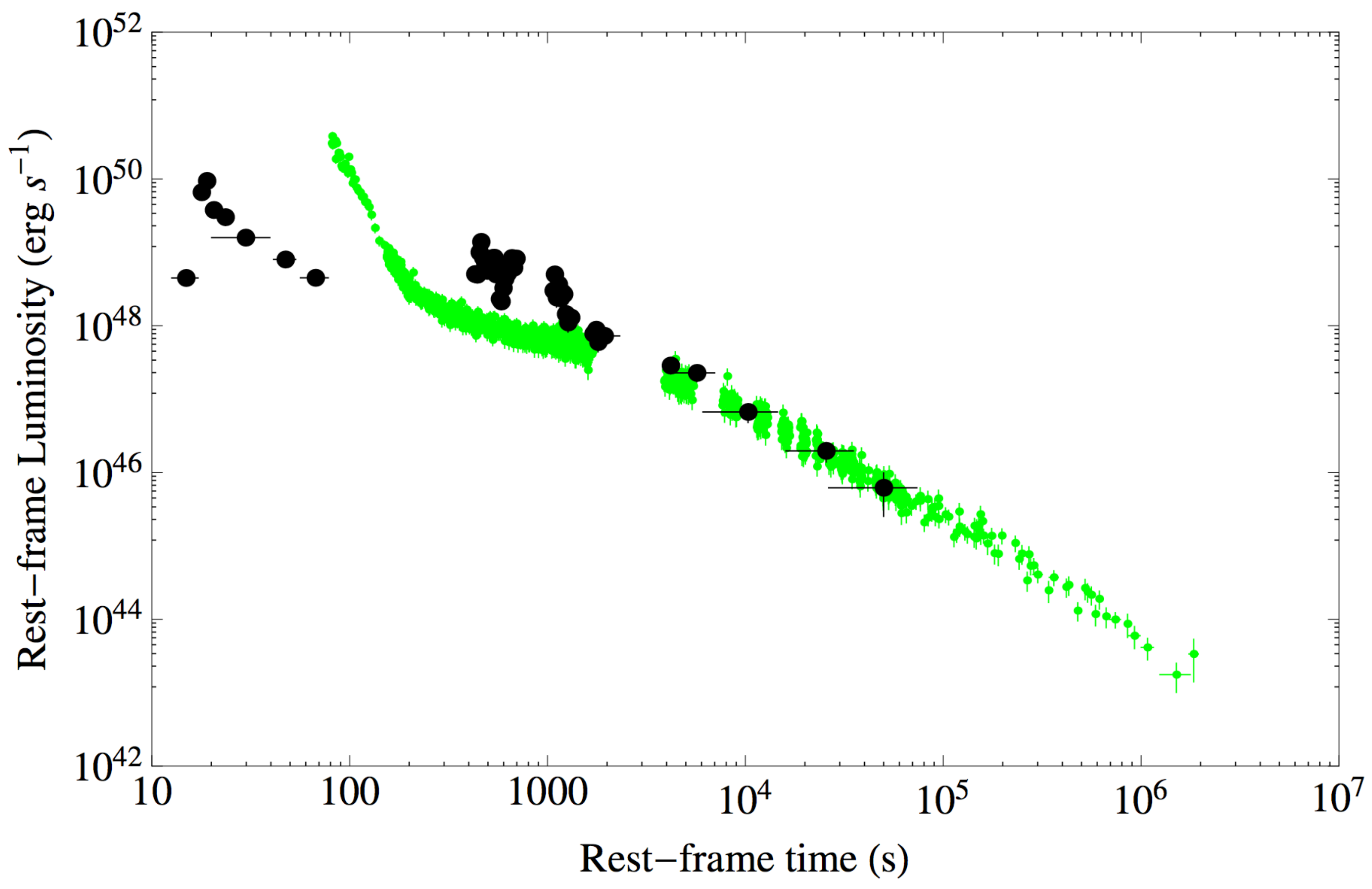}
\caption{X-ray luminosity of GRB 090423 (black points) compared with the one of GRB 090618 (green points), the prototype BdHN, by \citet{2014A&A...569A..39R}.}
\label{z8overlap}
\end{figure}

A crucial further step has been the identification as a BdHN of GRB 090423 \citep{2014A&A...569A..39R} at the extreme redshift of $z=8.2$ \citep{2009Natur.461.1258S,2009Natur.461.1254T}. On top of that, the LXRE of GRB 090423 overlaps perfectly with the ones of the GS (see Fig. \ref{z8overlap}), extending such a scaling law up to extreme cosmological distances. This result led to the necessity of checking such an universal behavior of the LXREs in BdHNe at redshifts larger than $z\sim1$ (see the sample list in Table \ref{EStable}).

It is clear by now that the afterglow analysis is much more articulated than previously expected and contains new specific signatures. When theoretically examined within our framework, these new signatures lead to specific information on the astrophysical nature of the progenitor systems \citep{2016arXiv160202732R}. In the present paper, we start by analysing the signatures contained in the LXREs at $t_{rf} \gtrsim 10^4$ s, where $t_{rf}$ is the rest-frame time after the initial GRB trigger.
In particular, we probe a further improvement for the existence of such an LXRE universal behavior of BdHNe by the introduction of a collimation correction.

In Section 2 we present an ``enlarged sample'' (ES) of 161 BdHNe observed up to the end of 2015.
In particular, we express for each BdHN: (1) redshift; (2) $E_{iso}$; and (3) the LXRE power law properties.
We probe the universality of the LXRE power law behavior as well as the absence of correlation with the prompt radiation phase of the GRB.
In Section 3 we introduce the collimation correction for the LXRE of BdHNe. This, in turn, will aim to the possible establishment of a new cosmological candle, up to $z \gtrsim 8$.
In Section 4 we present the inferences for the understading of the afterglow structure, and, in Section 5, we draw our conclusions.

\begin{figure*}
\centering
\includegraphics[width=0.48\hsize,clip]{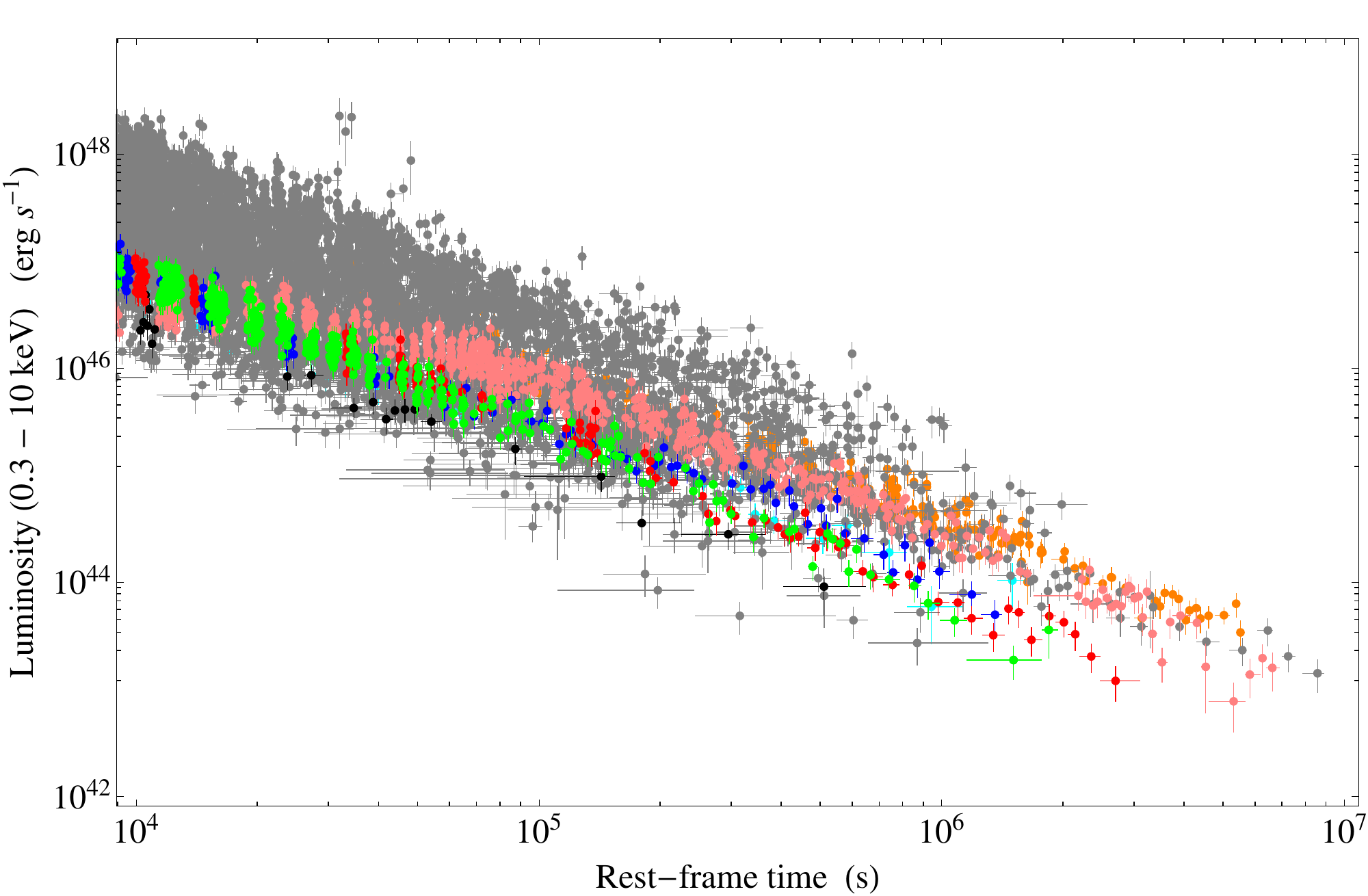}
\hfill
\includegraphics[width=0.49\hsize,clip]{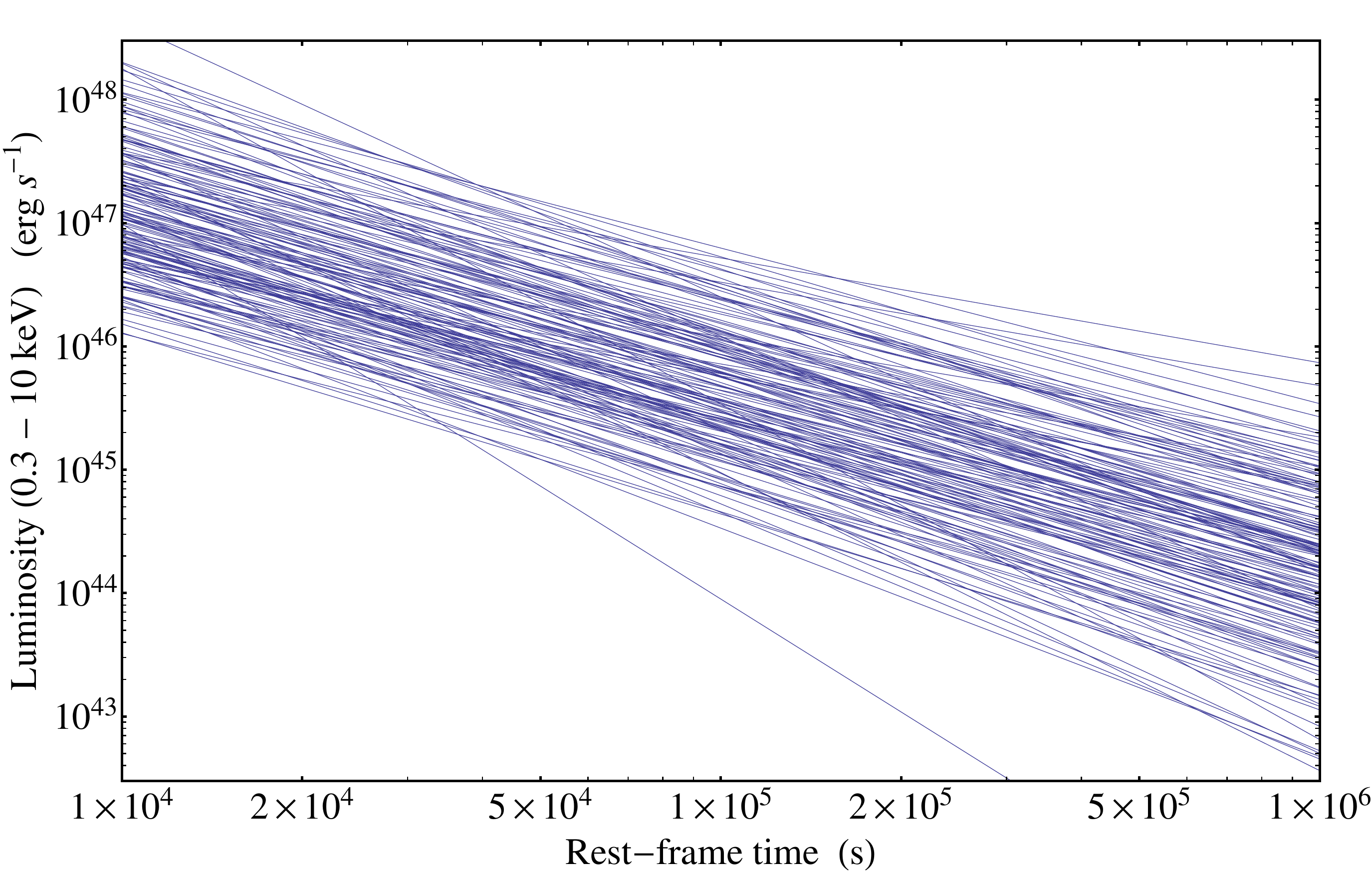}
\\
$\,$ \hfill (a) \hfill \hfill (b) \hfill  $\,$ \\
\caption{
Panel (a): LXRE luminosity light curves of all 161 sources of the ES (gray) compared with the ones of the GS: GRB 060729 (pink), GRB 061007 (black), GRB 080913B (blue), GRB 090618 (green), GRB 091127 (red), and GRB 111228 (cyan), plus GRB 130427A \citep[orange;][]{Pisani2013,2015ApJ...798...10R}.
Panel (b): power laws which best fit the luminosity light curves of the X-ray emissions of all 161 sources of the ES.
}
\label{descriptive}
\end{figure*}

\section{The BdHNe enlarged sample}

We have built a new sample of BdHNe, which we name ``enlarged sample'' (ES), under the following selection criteria:
\begin{itemize}
\item{measured redshift $z$;}
\item{GRB rest-frame duration larger than $2$ s;}
\item{isotropic energy $E_{iso}$ larger than $10^{52}$ erg; and}
\item{presence of associated \textit{Swift}/XRT data lasting at least up to $t_{rf}=10^4$ s.}
\end{itemize}

We collected $161$ sources, which satisfy our criteria, covering $11$ years of \textit{Swift}/XRT observations, up to the end of 2015, see Table \ref{EStable}.
The $E_{iso}$ of each source has been estimated using the measured redshift $z$ together with the best-fit parameters of the $\gamma$-ray spectrum published in the GCN circular archive\footnote{http://gcn.gsfc.nasa.gov/gcn3\_archive.html}.
The majority of the ES sources, 102 out of 161, have $\gamma$-ray data provided by \textit{Fermi}/GBM and Konus-WIND, which, with their typical energy bands being $10$--$1000$ keV and $20$--$2000$ keV, respectively, lead to a reliable estimate of the $E_{iso}$, computed in the ``bolometric'' $1$--$10^4$ keV band \citep{2001AJ....121.2879B}. The remaining sources of the ES have had their $\gamma$-ray emission observed by \textit{Swift}/BAT only, with the sole exception of one source observed by HETE.
The energy bands of these last two detectors, being $15$--$150$ keV and $8$--$400$ keV, respectively, lead to a standard estimate of $E_{iso}$ by extrapolation in the ``bolometric'' $1$--$10^4$ keV band \citep{2001AJ....121.2879B}.

\begin{figure*}
\centering
\includegraphics[width=0.47\hsize,clip]{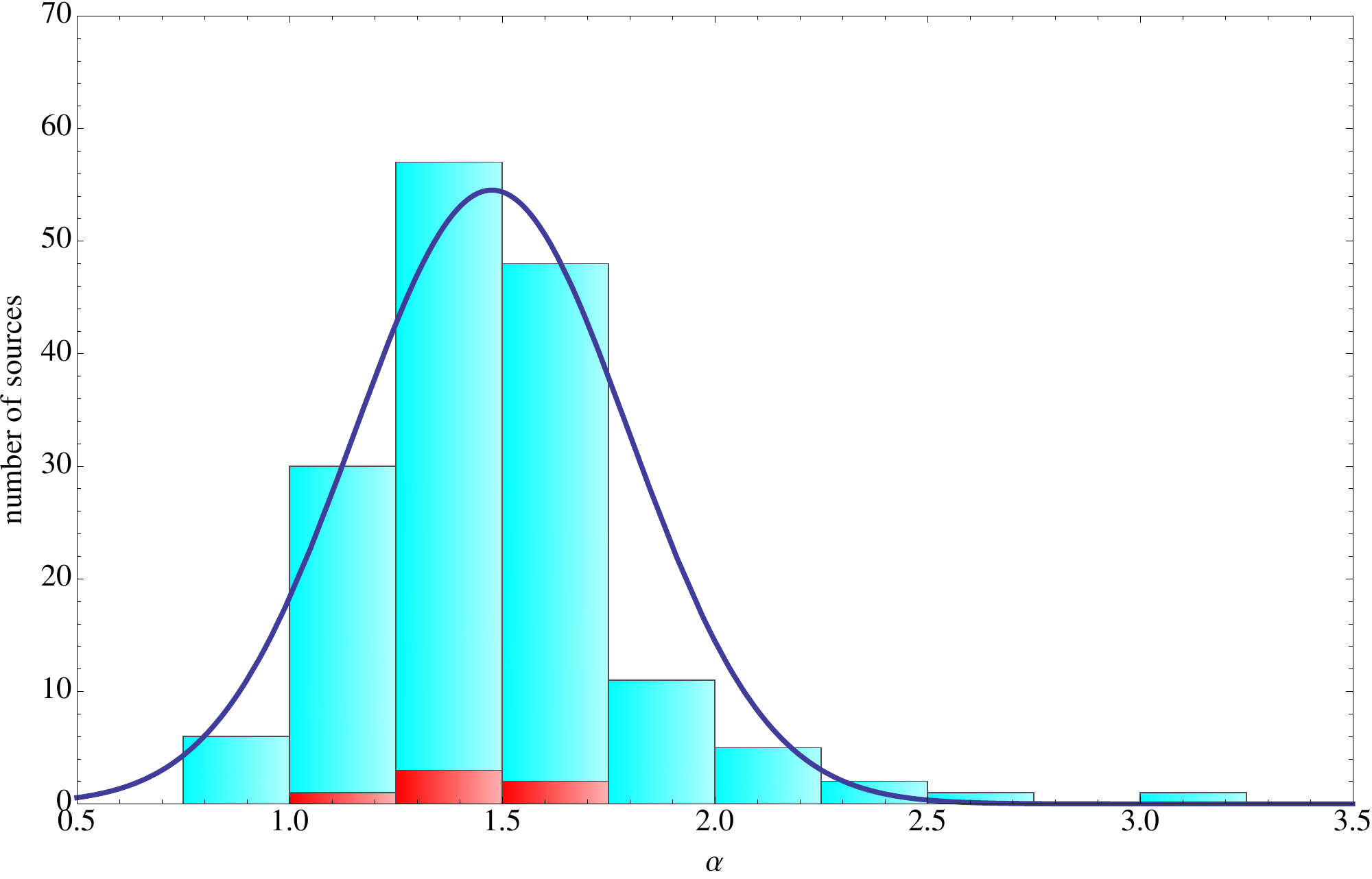}
\hfill
\includegraphics[width=0.47\hsize,clip]{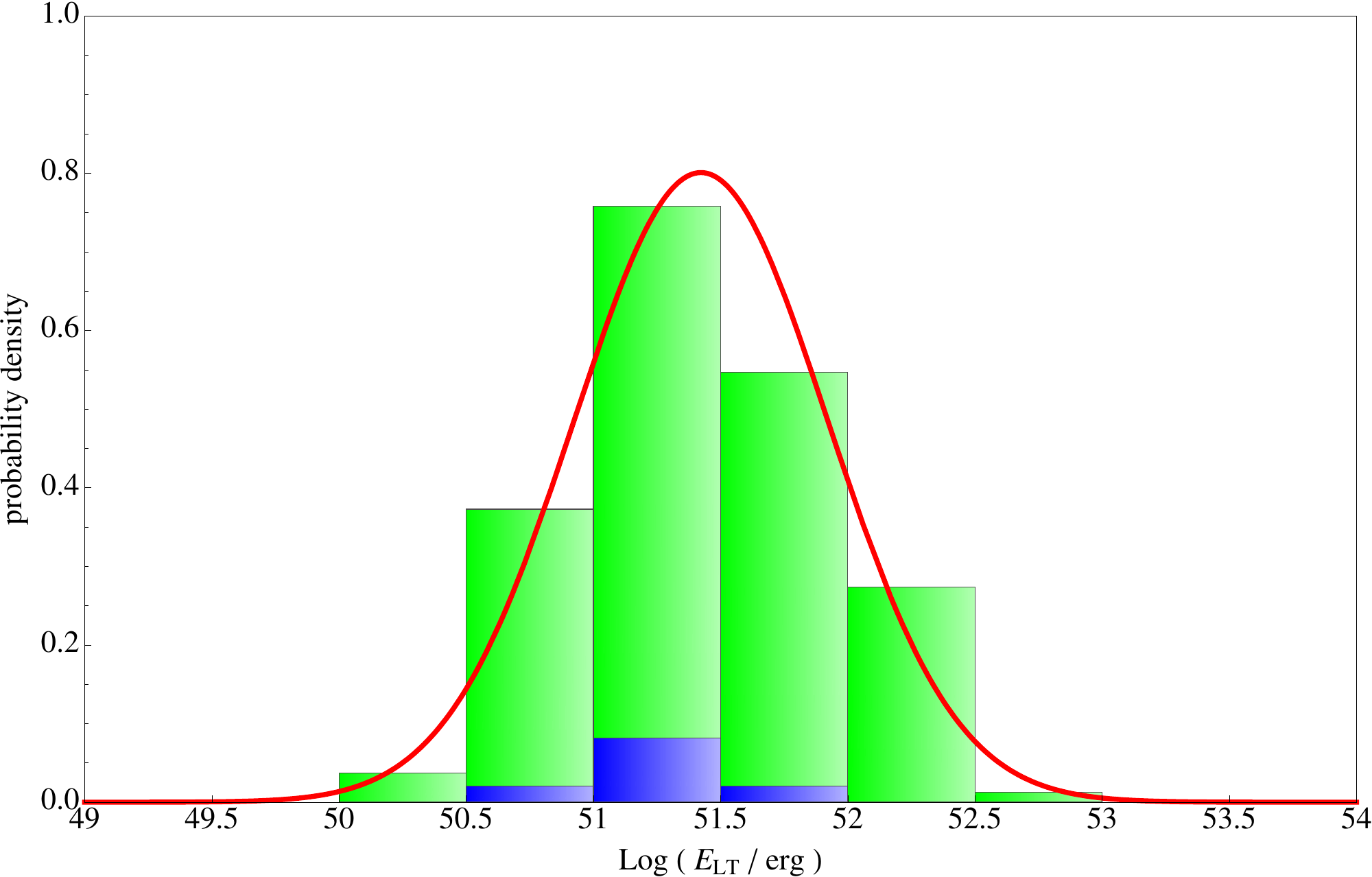}
\hfill \\
$\,$ \hfill (a) \hfill \hfill (b) \hfill  $\,$ \\ $\,$ \\
\includegraphics[width=0.47\hsize,clip]{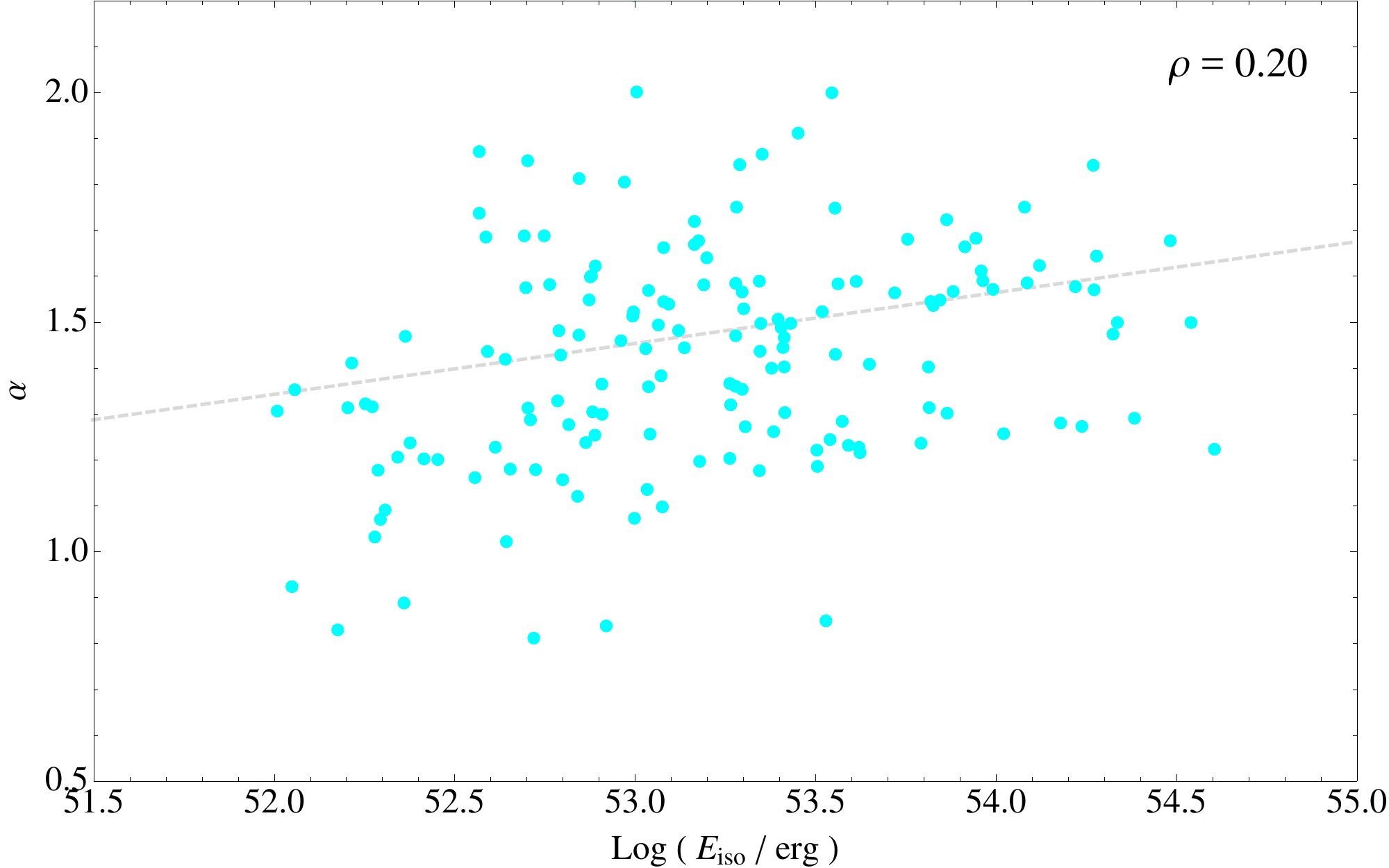}
\hfill
\includegraphics[width=0.47\hsize,clip]{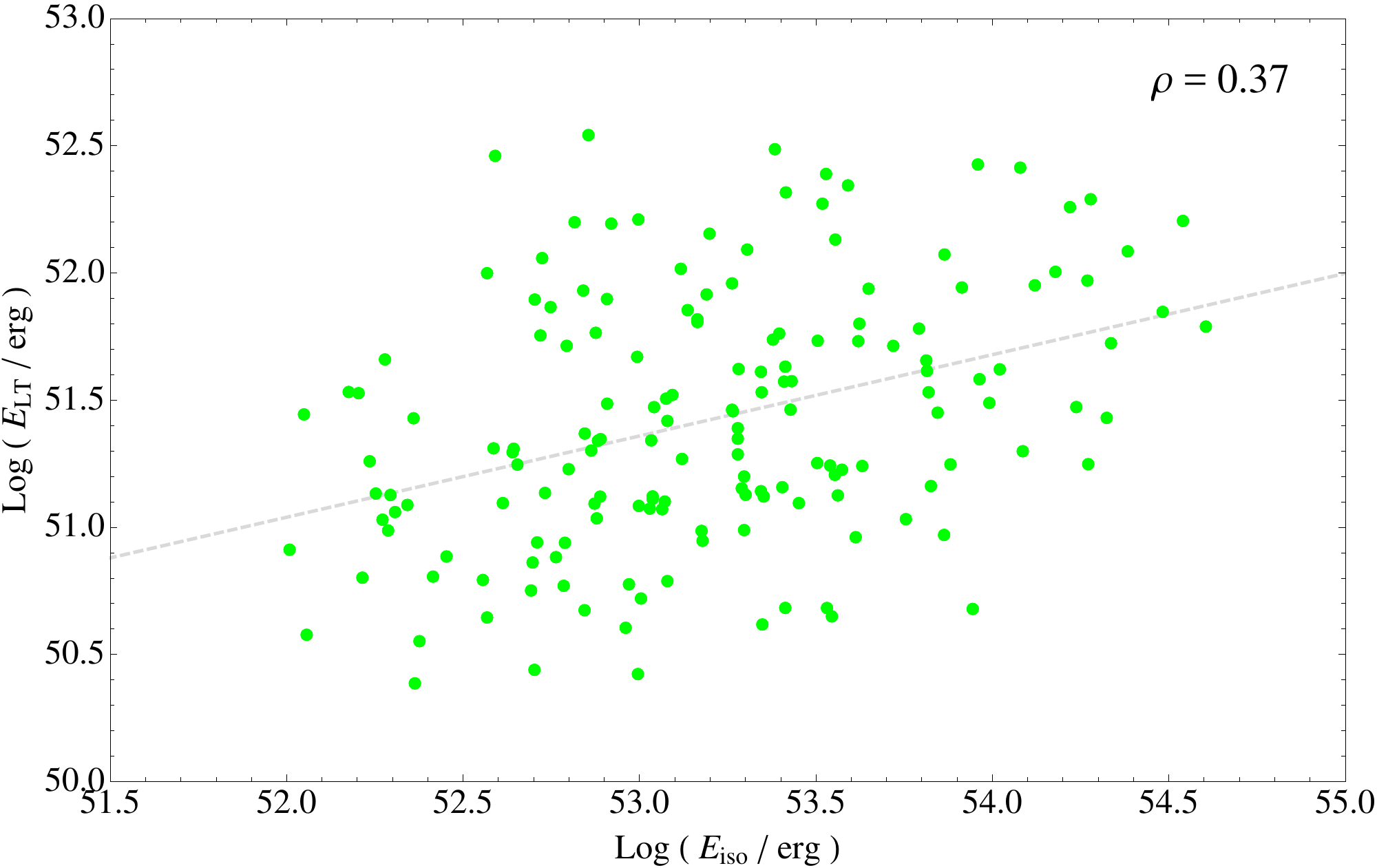}
\hfill \\
$\,$ \hfill (c) \hfill \hfill (d) \hfill  $\,$ \\
\caption{
Panel (a): distribution of the LXRE power law indexes $\alpha$ within the ES (cyan) compared to the one of the GS (red). Such a distribution follows a Gaussian behavior (blue line) with a mean value of $\mu_{\alpha}=1.48$ and a standard deviation of $\sigma_{\alpha}=0.32$.
Panel (b): probability distribution of the LXRE integrated energies within the time interval $10^4$--$10^6$ s in the rest-frame after the initial GRB trigger for all the sources of the ES (in green) compared with the GS (in blue). The solid red line represents the Gaussian function which best fits the ES data in logarithmic scale. Its mean value is $\mu_{\mathrm{Log}_{10}(E_{LT})}=51.40$, while its standard deviation is $\sigma_{\mathrm{Log}_{10}(E_{LT})}=0.47$.
Panel (c): scatter plot of $\alpha$ versus $E_{iso}$ (cyan points) in logaritcmic scale. The dashed gray line is the best linear fit on such points. If we look at the correlation coefficient of these data points, $\rho=0.20$, we conclude that there is no evidence for correlation between the two quantities.
Panel (d): scatter plot of $E_{LT}$ versus $E_{iso}$ (green points) in logaritcmic scale. The dashed gray line is the best linear fit on such points. If we look at the correlation coefficient of these data points, $\rho=0.37$, we conclude that there is no evidence for correlation between the two quantities.
}
\label{descriptive2}
\end{figure*}

We compare the \textit{Swift}/XRT isotropic luminosity light curve $L^{iso}_{rf}$ for 161 GRBs of the ES in the common rest-frame energy range of $0.3\,$--$\,10$ keV. We initially convert the observed \textit{Swift}/XRT flux $f_{obs}$ as if it had been observed in the $0.3\,$--$\,10$ keV rest-frame energy range. In the detector frame, the $0.3\,$--$\,10$ keV rest-frame energy range becomes $[0.3/(1+z)]\,$--$\,[10/(1+z)]$ keV, where $z$ is the redshift of the GRB. We assume a simple power law function as the best fit for the spectral energy distribution of the \textit{Swift}/XRT data\footnote{http://www.swift.ac.uk/}:
\begin{equation}
\frac{dN}{dA\,dt\,dE} \propto E^{-\gamma}\,.
\label{spettro_pl}
\end{equation}
Therefore, we can compute the flux light curve in the $0.3\,$--$\,10$ keV rest-frame energy range, $f_{rf}$, multiplying the observed one, $f_{obs}$, by the k-correction factor:
\begin{equation}
f_{rf} = f_{obs} \frac{\int_{\frac{0.3\,keV}{1+z}}^{\frac{10\,keV}{1+z}}E^{1-\gamma}dE}{\int_{0.3\,keV}^{10\,keV}E^{1-\gamma}dE} = f_{obs} (1+z)^{\gamma-2}\,.
\label{flusso_1}
\end{equation}
Then, to compute the isotropic X-ray luminosity $L_{iso}$, we have to multiply $f_{rf}$ by the spherical surface having the luminosity distance as radius
\begin{equation}
L_{iso} = 4 \, \pi \, d_l^2(z) f_{rf}\,,
\label{luminosity}
\end{equation}
where we assume a standard cosmological $\Lambda$CDM model with $\Omega_m = 0.27$ and $\Omega_{\Lambda}=0.73$. Finally, we convert the observed times into rest-frame times $t_{rf}$:
\begin{equation}
\label{time_correction}
t_{rf} = \frac{t_{obs}}{1+z}\,.
\end{equation}

We then fit the whole isotropic luminosity light-curve late phase with a decaying power law function defined as follows:
\begin{equation}
L_{iso}(t_{rf}) = L_0 \,\, t_{rf}^{\,\,\,-\alpha} \, ,
\label{powerlaw}
\end{equation}
where $\alpha$, the power law index, is a positive number, and $L_0$ is the luminosity at an arbitrary time $t_{rf}=t_0$ after the GRB trigger in the rest-frame of the source. All the power laws are shown in Fig. \ref{descriptive}b.
Fig. \ref{descriptive2}a shows the distribution of the $\alpha$ indexes within the ES. Such a distribution follows a Gaussian behavior with a mean value of $\mu_{\alpha}=1.48$ and a standard deviation of $\sigma_{\alpha}=0.32$.
The LXRE luminosity light curves of the ES in the $0.3$--$10$ keV rest-frame energy band are plotted in Fig. \ref{descriptive}a, compared to the curves of the GS. Fig. \ref{descriptive}a shows that the power laws within the ES span around two orders of magnitude in luminosity.
The spread of the LXRE light curves in the ES is better displayed by Fig. \ref{descriptive2}b which shows the distribution within the ES of the LXRE integrated energies $E_{LT}$ defined as:
\begin{equation}
E_{LT} \equiv \int_{10^4 s}^{10^6 s} \! {L_{iso}(t_{rf}) \, \mathrm{d}t_{rf}} \,.
\end{equation}

We choose to represent the spread of the LXRE luminosity light curves with the late integrated energy $E_{LT}$ at late times ($t_{rf}=10^4$--$10^6$ s) instead of the luminosity $L_{iso}$ at a particular time for two reasons: (1) there is no evidence for a particular time in which to compute $L_{iso}$ and (2) we want to have a measurement of the spread as much as possible independent from the slopes, whose dispersion causes the mixing of part of the light curves over time (see Fig. \ref{descriptive}b).
The integration time interval $t_{rf}=10^4$--$10^6$ reasonably contains most of the data in the late power law behavior. In fact, the lower limit $t_{rf}=10^4$ s is basically the average of the initial time for our linear fits on the data (more precisely $t^{start}_{rf}=9167.13$ s), while the upper limit $t_{rf}=10^6$ has been chosen because only $14\%$ of the ES have X-ray data over such rest-frame time.

The solid red line in Fig. \ref{descriptive2}b represents the Gaussian function that best fits the late integrated energies $E_{LT}$ in logarithmic scale. Its mean value is $\mu_{\mathrm{Log}_{10}(E_{LT})}=51.40$, while its standard deviation is $\sigma_{\mathrm{Log}_{10}(E_{LT})}=0.47$.

The LXRE power-law spread, given roughly by $2\sigma_{\mathrm{Log}_{10}(E_{LT})}=0.94$, is larger in respect to the previous work of \citet{Pisani2013}, which results as $2\sigma_{\mathrm{Log}_{10}(E_{LT})}=0.56$. This is clearly due to the significant growth of the number of BdHNe composing the ES (161) in respect to the ones of the GS (6).

Furthermore, Fig. \ref{descriptive2}c-d show the scatter plots of the values of $\alpha$ and $E_{LT}$, respectively, versus the values of the $E_{iso}$ for all the sources of the ES. In both cases the correlation factors, $\rho=0.20$ and $\rho=0.37$ respectively, are low, confirming that there is no evidence for a correlation between the LXRE power law behavior and the isotropic energy emitted by the source during the prompt radiation.

These results address a different aspect than the ones by \citet{2013MNRAS.428..729M}. There, the authors, after correctly noticing the difficulties of the traditional afterglow model \citep{MeszarosRees1997,Sari1998b}, attempt to find a model-independent correlation between the X-ray light curve observed in both short and long GRBs with their prompt emission.
In their work, \citet{2013MNRAS.428..729M} have considered the integrated X-ray emission over the entire light curve observed by XRT, following $\sim 300$ s after the GRB trigger both for short and long GRBs.
Such an emission is clearly dominated by the contribution at $t_{rf} < 10^4$ s, where a dependence from the $E_{iso}$ is self-evident from the above Fig. \ref{scaling} and Fig. \ref{nesting}.
Our approach instead solely applies to the BdHNe: (1) long GRBs, and (2) $E_{iso} > 10^{52}$ erg. This restricts the possible sources in the \citet{2013MNRAS.428..729M} sample to $70$ GRBs: in the present article, we consider a larger sample of $161$ BdHNe. Moreover, (3) our temporal window starts at $t_{rf} \gtrsim 10^4$ s. Under these three conditions, our result of the universal LXRE behavior has been found.

\section{Collimation}

\begin{figure*}
\centering
\includegraphics[width=0.48\hsize,clip]{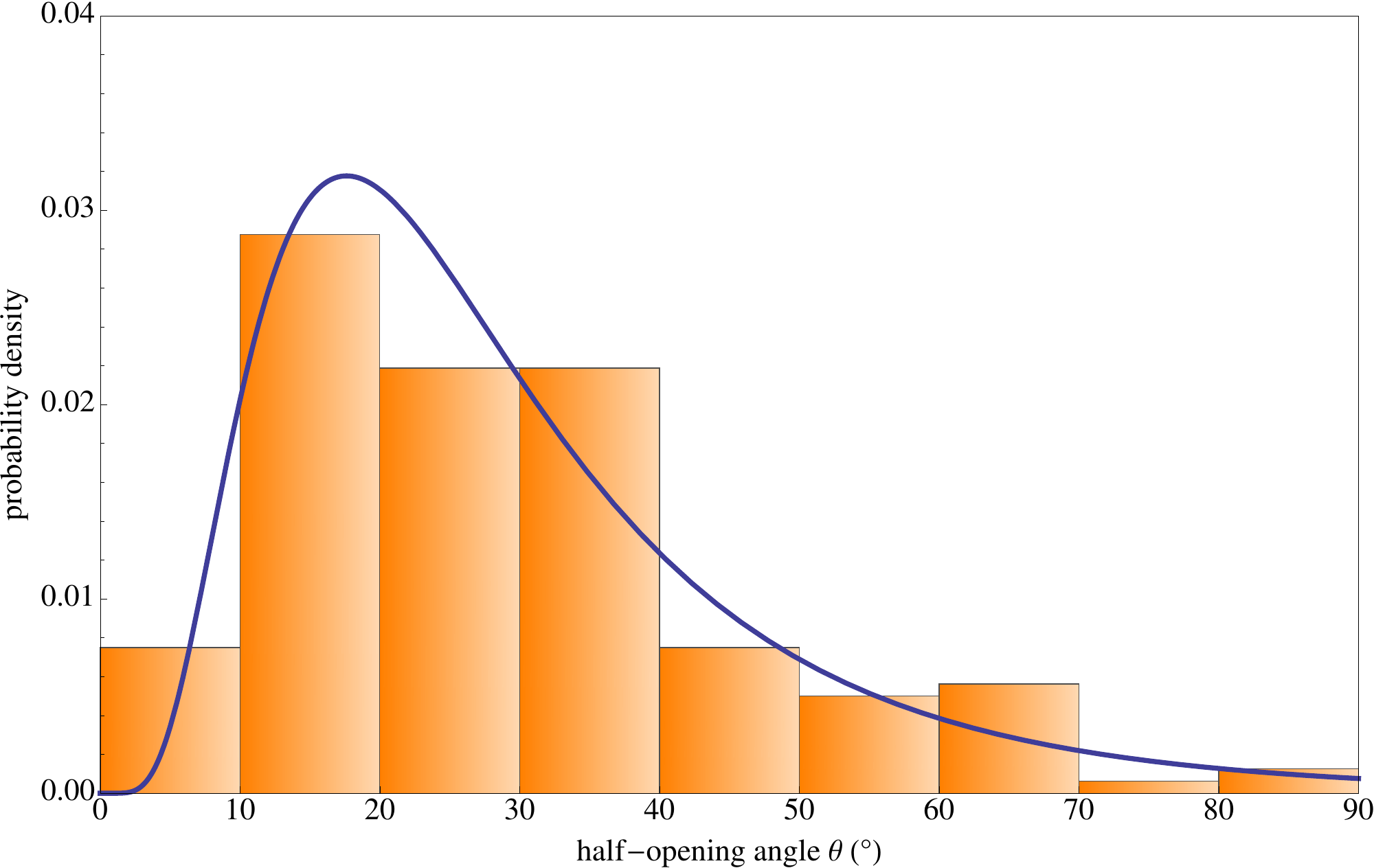}
\hfill
\includegraphics[width=0.48\hsize,clip]{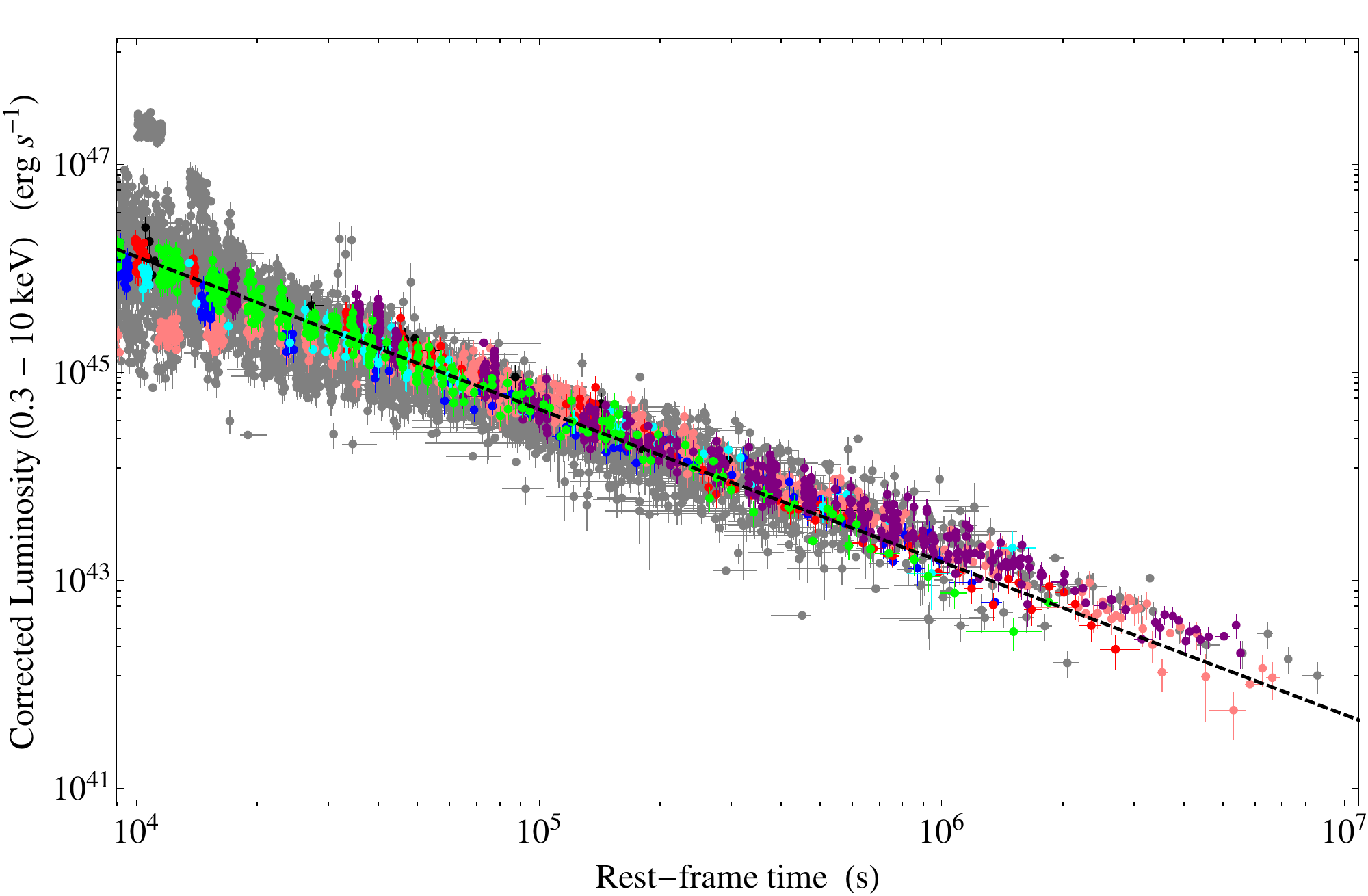}
\\
$\,$ \hfill (a) \hfill \hfill (b) \hfill  $\,$ \\
\caption{
Left panel (a): probability distribution of the half-opening angle $\theta$ within the ES. The blue solid line represents a logarithmic normal distribution, which best fits the data. This distribution has a mode of $Mo_{\theta}=17.62^\circ$, a mean of $\mu_{\theta}=30.05^\circ$, a median of $Me_{\theta}=25.15^\circ$, and a standard deviation of $\sigma_{\theta}=19.65^\circ$.
Right panel (b): corrected LXRE luminosity light curves of all 161 sources of the ES (gray) compared to the ones of the GS: GRB 060729 (pink), GRB 061007 (black), GRB 080913B (blue), GRB 090618 (green), GRB 091127 (red), and GRB 111228 (cyan), plus GRB 130427A \citep[purple;][]{Pisani2013,2015ApJ...798...10R}. The black dotted line represents the universal LXRE power law, namely the linear fit of the late emission of GRB 050525A.}
\label{angles}
\end{figure*}

We here propose to reduce the spread of the LXRE power laws within the ES by introducing a collimation effect in the emission process. In fact, if the emission is not isotropic, Fig. \ref{descriptive}a-b should actually show overestimations of the intrinsic LXRE luminosities.
By introducing a collimation effect, namely assuming that the LXREs are not emitted isotropically but within a double-cone region having half-opening angle $\theta$, we can convert the isotropic $L_{iso}(t_{rf})$ to the intrinsic LXRE luminosity $L^{intr}(t_{rf})$ as:
\begin{equation}
L^{intr}(t_{rf}) = L_{iso}(t_{rf}) \, (1 - \cos \theta) \, .
\label{Lumintrinsec}
\end{equation}

From Eq. \ref{Lumintrinsec}, an angle $\theta$ can be inferred for each source of the ES if an intrinsec universal LXRE light curve $L^{intr}(t_{rf})$ is assumed.
For example, assuming an intrinsic standard luminosity $L^{intr}_{0}$, at an arbitrary time $t_{rf}=t_0$, Eq. \ref{Lumintrinsec} becomes $L^{intr}_{0} = L_{0} \, (1 - \cos \theta)$, which, in principle, could be used to infer $\theta$ for each source. 
On the other hand, for the same reasons expressed in the previous section, we choose to estimate the angles $\theta$ using the LXRE integrated energy $E_{LT}$ instead of the $L_{iso}$ at a particular time.
Therefore, we simply integrate Eq. \ref{Lumintrinsec} in the rest-frame time interval $10^4$--$10^6$ s:
\begin{equation}
\int_{10^4 s}^{10^6 s} \! {L_{intr}(t_{rf}) \, \mathrm{d}t_{rf}} = \int_{10^4 s}^{10^6 s} \! {L_{iso}(t_{rf}) \, (1 - \cos \theta) \, \mathrm{d}t_{rf}} \,,
\end{equation}
obtaining, consequentially
\begin{equation}
E^{intr}_{LT} = E_{LT} \, (1 - \cos \theta) \, .
\label{Eint}
\end{equation}

By assuming a universal $E^{intr}_{LT}$ for all BdHNe, it is possible to infer $\theta$ for each source of the ES.
We assume GRB 050525A, having the lowest $E_{LT}$ within the ES, as our unique ``isotropic'' source, namely, in which we can impose $E^{intr}_{LT} = E_{LT}$, which automatically gives $\theta = 90^\circ$, which means that the LXRE luminosity is emitted over all the isotropic sphere.
On the top of having the weakest LXRE over 11 years of \textit{Swift}/XRT observations, GRB 050525A: \textit{(a)} has been observed by Konus-WIND in the $\gamma$-rays \citep{2005GCN..3474....1G}, then its $E_{iso}$ estimate is reliable; \textit{(b)} has a reliable late X-ray slope given by a complete \textit{Swift}/XRT light curve (showing a late power law behavior from $4000$ to $7\times10^5$ s in the rest frame); \textit{(c)} has an associated supernova \citep{2006ApJ...642L.103D,2006IAUC.8696....1D}.
An instrumental selection effect cannot affect this choice since \textit{Swift}/XRT can easily detect and follow X-ray emissions weaker than that of GRB 050525A. Some examples are the two X-ray luminosity light curves shown of GRB 060218 and GRB 101219B in Fig. \ref{XRF}-a.
Furthermore, in the case of a future observation of a BdHN showing a $E_{LT}$ weaker than the one of GRB 050525A, a renormalization of the angle distribution of Fig. \ref{angles}a down to lower angle values with respect to the new minimum one would be required, leaving the overall angle distribution unaltered.
All of these facts make GRB 050525A a robust ``isotropic'' BdHN candidate.
With $E^{intr}_{LT} = E_{LT}^{050525A} = 2.43 \times 10^{50}$ now fixed, a half-opening angle $\theta$ is inferred for each source of the ES using Eq. \ref{Eint}.
The values of $\theta$ are listed in Table \ref{EStable}.
Fig. \ref{angles}a shows the probability distribution of the half-opening angle $\theta$ within the ES. The blue solid line represents a logarithmic normal distribution, which best fits the data. This distribution has a mode of $Mo_{\theta}=17.62^\circ$, a mean of $\mu_{\theta}=30.05^\circ$, a median of $Me_{\theta}=25.15^\circ$, and a standard deviation of $\sigma_{\theta}=19.65^\circ$.
Moreover, it is possible to verify that, by correcting the $L^{iso}_{rf}$ light curve of each ES source for its corresponding $\theta$, an overlap of the LXRE luminosity light curves as good as the one seen in the GS by \citet{Pisani2013} shown in Fig. \ref{scaling} is obtained.
Since the LXRE follows a power law behavior, we can quantify the tightness of the LXREs overlap looking at the correlation coefficient $\rho$ between all the luminosity light-curve data points of the ES sources in log-log scale. Considering the data points of the LXRE power laws within the $10^4$--$10^6$ s time interval (where we have defined $E_{LT}$), we obtain $\rho=-0.94$ for the GS, $\rho=-0.84$ for the ES before the collimation correction, and $\rho=-0.97$ after the correction.
Therefore, the collimation correction not only reduces the spread of the LXREs within the ES, but makes the LXREs overlap even tighter than the one previously found in the GS.

Finally, in order to test the robustness of our results, we do the same analysis excluding, by the ES, the sources seen only by \textit{Swift}/BAT or HETE and not by \textit{Fermi}/GBM or Konus-WIND, under the hypothesis that their $E_{iso}$ estimates are unreliable. The results obtained using this new sample called ES2, namely the typical value and the dispersion of $\alpha$, $E_{LT}$, and $\theta$ are summarized in Table \ref{tabESES2}.
There is no significant difference between the results obtained from the two samples. Therefore, we conclude that a possible wrong estimate of $E_{iso}$ for the sources observed by only \textit{Swift}/BAT or HETE and not by \textit{Fermi}/GBM or Konus-WIND does not bias our results.

\begin{table}
\caption{Summary of the results of this work obtained by the complete ES sample in comparison with the ones arising using the sample ES2, namely the ES deprived by the sources seen by \textit{Swift}/BAT or HETE only.}
\centering
\begin{tabular}{lcc}
\hline\hline
Sample                                                                & ES                          & ES2                         \\
\hline
Sources Number                                                 & $161$                     & $102$                      \\
$\alpha$                                                              & $1.48 \pm 0.32$     & $1.45 \pm 0.24$      \\
$\mathrm{Log}_{10}(E_{LT}/\mathrm{erg)}$       & $51.40 \pm 0.47$   & $51.47 \pm 0.48$    \\
$\theta$ ($^\circ$)                                               & $30.05 \pm 19.65$ & $28.26 \pm 17.85$  \\
$\theta$ mode ($^\circ$)                                     & $17.62$                  & $17.08$                    \\
$\theta$ median ($^\circ$)                                  & $25.15$                  & $23.90$                    \\
\hline
\end{tabular}
\label{tabESES2}
\end{table}

\section{Inferences for the understanding of the X-ray afterglow structure}

In the last 25 years we have seen in the GRB community a dominance of the fireball model, which sees the GRB as a single astrophysical system, the ``Collapsar'', originating from an ultra-relativistic jetted emission described by the synchrotron/self-synchrotron Compton (SSC) and traditional afterglow models
\citep[see, e.g. ][ and references therein]{1993ApJ...405..273W, 1992MNRAS.258P..41R, RevModPhys.76.1143, 2009ARA&A..47..567G, 2015PhR...561....1K}.
Such methods have been systematically adopted to different types of GRBs like, for example, the short, hard GRB 090510 \citep{2010ApJ...716.1178A}, the high energetic long GRB 130427A \citep{2014ApJ...781...37P}, the low energetic short GRB 051221A \citep{2006ApJ...650..261S}, and the low energetic long GRB 060218 \citep{2006Natur.442.1008C,2006Natur.442.1014S}, independently from the nature of their progenitors.

In the recent four years, substantial differences among seven distinct kinds of GRBs have been indicated, presenting different spectral and photometrical properties on different time scales \citep{2016arXiv160202732R}. The discovery of several long GRBs showing multiple components and evidencing the presence of a precise sequence of different astrophysical processes during the GRB phenomenon \citep[e.g. ][]{Izzo2012,Penacchioni2012}, led to the introduction of a novel paradigm expliciting the role of binary sources as progenitors of the long GRB-SN connection.
This has led to the formulation of the IGC paradigm \citep{2001ApJ...555L.113R, Ruffini2007b, Izzo2012b, Rueda2012, 2014ApJ...793L..36F, 2015ApJ...798...10R}.
Within the IGC paradigm, a tight binary system composed of a carbon-oxygen core (CO$_{\mathrm{core}}$) undergoing a supernova (SN) explosion in the presence of a binary NS companion has been suggested as the progenitor for long gamma-ray bursts.
Different scenarios occur depending on the distance between the CO$_{\mathrm{core}}$ and the NS binary companion \citep{2015ApJ...812..100B}. Correspondingly two different sub-classes of long bursts have been shown to exist \citep[for details, see ][]{2015ApJ...798...10R, 2016arXiv160202732R}.
A first long burst sub-class occurs when the CO$_{\mathrm{core}}$--NS binary separation $a$ is so large (typically $a > 10^{11}$ cm, see, e.g., \citet{2015ApJ...812..100B} that the accretion of the SN ejecta onto the NS is not sufficient to have the NS reach its critical mass, $M_{crit}$, for gravitational collapse to a BH to occur. The hypercritical accretion of the SN ejecta onto the NS binary companion occurs in this case at rates below $10^{-2}$ solar masses per second and is characterized by a large associated neutrino emission \citep{1972SvA....16..209Z, 1973PhRvL..31.1362R, Rueda2012, 2014ApJ...793L..36F}.
We refer to such systems as X-ray flashes (XRFs).
A second long burst sub-class occurs when the CO$_{\mathrm{core}}$--NS binary is more tightly bound ($a < 10^{11}$ cm, see, e.g., \citet{2015ApJ...812..100B}. The larger accretion rate of the SN ejecta, e.g., $10^{-2}$--$10^{-1}$ solar masses per second, leads the companion NS to easily reach its critical mass $M_{crit}$ \citep{Rueda2012, 2014ApJ...793L..36F, 2015ApJ...812..100B}, leading to the formation of a BH.
We refer to such systems as binary-driven hypernovae \citep[BdHNe, see, e.g., ][]{ 2014A&A...565L..10R, 2015ApJ...798...10R}.
A main observational feature, which allows us to differentiate BdHNe from XRFs is the isotropic $\gamma$-ray energy $E_{iso}$ being larger than $10^{52}$ erg. Such a separation energy value is intimately linked to the binary separation $a$ of the binary progenitor and the consequent birth or not of the BH \citep[for details, see][]{2016arXiv160202732R}.

\begin{figure}
\centering
\includegraphics[width=\hsize,clip]{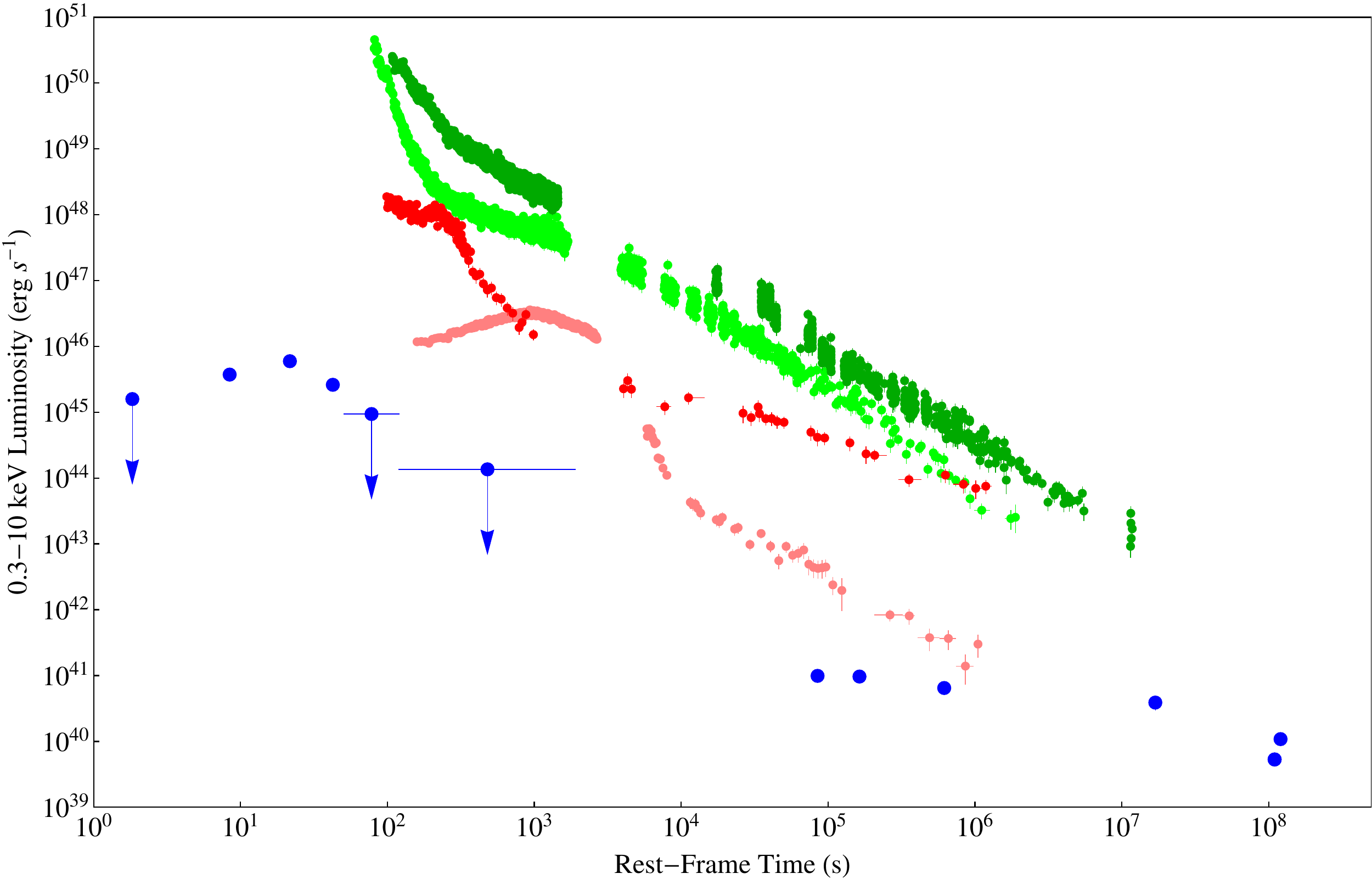} \\
\caption{Comparison between rest-frame luminosity light curves of proto-tipycal BdHNe and XRFs sources. The BdHNe shown are GRB 130427A (dark green) and GRB 090618 (light green); while the XRF shown are GRB 101219B (red), GRB 060218 (pink), and GRB 980425 (blue).}
\label{XRF}
\end{figure}

Thanks to the XRT instrument on board the \textit{Swift} satellite \citep{2004ApJ...611.1005G, 2007A&A...469..379E, 2010A&A...519A.102E}, we can compare and contrast the X-ray afterglow emissions of BdHNe and XRFs (see Fig. \ref{XRF}).
The typical X-ray afterglows of XRFs can be divided into two main parts: an initial bump with rapid decay, followed by an emerging slight decaying power law (see Fig. \ref{XRF}).
The typical X-ray afterglow light curve of BdHNe can be divided into three different parts: (1) an initial spike followed by an early steep power law decay; (2) a plateau phase; and (3) a late power law decay, the LXRE, which is the presented in this work \citep{2015ApJ...798...10R,2006ApJ...642..389N}.
The treatment of the first parts of the X-ray afterglow of BdHNe, namely the spike, the initial steep decay, and the plateau phase, indeed fundamental within the BdHN picture, is beyond the scope of this article and it will be extensively shown in forthcoming works \citep[][both in preparation]{Spikesinprep,Plateauinprep}.

The universalities of the LXRE outlined in this article are then explained within the IGC paradigm, originating from the interaction of the GRB with the SN ejecta.
The constancy of the late power law luminosity in the rest frame is now explained in terms of the constancy in mass of the SN ejecta, which is standard in a BdHN \citep{Rueda2012, 2014ApJ...793L..36F, 2015ApJ...812..100B}.

It is appropriate to point out that no achromatic ``Jet-break'' effect has been observed in any of the 161 sources of our ES. We recall that the achromatic ``Jet-break'' effect is a consequence of relativistic jet pictures (Lorentz factor $\Gamma \sim 100$--$200$), in which a change of slope is expected in the late X-ray light curve \citep[see e.g.][ and references therein]{1993ApJ...405..273W, 1992MNRAS.258P..41R, RevModPhys.76.1143, 2009ARA&A..47..567G, 2015PhR...561....1K}, which clearly does not apply in the case of the BdHN following the IGC model.
In this scenario, a velocity of expansion $v \sim 0.8 c$ (Lorentz factor $\Gamma \sim 2$) is found, indicating that the collimation of the SN ejecta originates in a mildly relativistic regime \citep{2014A&A...565L..10R, 2015ApJ...798...10R}.
This cannot be related to the ultra-relativistic jet emission recalled above, considered in the early work of \citet{2001ApJ...562L..55F} and continued all the way to the more recent results presented by \citet{2013MNRAS.428.1410G}.
These authors attempted to explain all GRBs as originating from a single object with an intrinsic energy approximately of $10^{50}$ erg \citep{2001ApJ...562L..55F} or $10^{48}$ erg \citep{2013MNRAS.428.1410G}: the different energetics and structures of all the GRBs were intended to be explained by the beaming effect with different ultra-relativistic Lorentz factors $\Gamma \sim 100$--$200$.
Indeed, it is by now clear \citep[see ][]{2015ApJ...798...10R, 2016arXiv160202732R} that at least seven different classes of GRBs exist, each with different progenitors, different energies, and different spectra. In no way these distinct classes can be explained by a single common progenitor, using simply relativistic beaming effects.

\section{Conclusions}

In this work, we give new statistical evidence for the existence of a universal behavior for the LXREs of BdHNe, introducing the presence of a collimation effect in such emission, and presenting the common LXRE energy $E^{intr}_{LT} = E_{LT}^{050525A} = 2.43 \times 10^{50}$ as a standard candle.
We build an ``enlarged sample'' (ES) of 161 BdHNe, and focus on their LXREs and then we introduce a collimation effect.
These analyses lead us to the following results.

\textbf{1)} We find for the ES an increased variability in the decaying LXRE power law behavior in respect to the result previously deduced by \citet{Pisani2013}. The typical slope of the power law characterizing the LXRE is $\alpha=1.48 \pm 0.32$ (GS: $\alpha=1.44 \pm 0.18$), while the late-time integrated luminosity between $10^4$--$10^6$ seconds in the rest frame is $\mathrm{Log}_{10}(E_{LT}/\mathrm{erg)}=51.40 \pm 0.47$ (GS: $\mathrm{Log}_{10}(E_{LT}/\mathrm{erg)}=51.15 \pm 0.28$).

\textbf{2)} The introduction of a collimation in the LXRE recovers a universal behavior.
Assuming a double-cone shape for the LXRE region, we obtain a distribution of half-opening angles peaking at $\theta=17.62^\circ$, with a mean value of $30.05^\circ$, and a standard deviation of $19.65^\circ$, see Fig. \ref{angles}a. 

\textbf{3)} The application of the collimation effect to the LXREs of the ES indeed reduces the scattering of the power law behavior found under the common assumption of isotropy; see Fig. \ref{descriptive}a-b. The power law scattering of the LXREs, after being corrected by the collimation factor, results in being even lower than the one found in the GS; see Fig. \ref{angles}b.

The fact that these extreme conditions neither were conceived nor are explained within the traditional ultra-relativistic jetted SSC model \citep[see, e.g. ][ and references therein]{1993ApJ...405..273W, 1992MNRAS.258P..41R, RevModPhys.76.1143, 2009ARA&A..47..567G, 2015PhR...561....1K}, in view also of the clear success of the IGC paradigm in explaining the above features, comes as a clear support to a model for GRBs strongly influenced by the binary nature of their progenitors, involving a definite succession of selected astrophysical processes for a complete description of the BdHNe.

These intrinsic signatures in the LXREs of BdHNe, independent from the energetics of the GRB prompt emission, open the perspective for a standard candle up to $z \gtrsim 8$.

It is remarkable that the universal behavior occurs in the rest-frame time interval $10^4$--$10^6$ s, which precisely corresponds to the temporal window of the early observations of Beppo-SAX at the time of the afterglow discovery (see Fig \ref{BeppoSAX}).

\acknowledgments

We thank both the editor and the referee for the fruitful correspondence that improved the presentation of our results.
This work made use of data supplied by the UK Swift Science Data Center at the University of Leicester.
J.~A.~R. acknowledges the support by the International Cooperation Program CAPES-ICRANet financed by CAPES-Brazilian Federal Agency for Support and Evaluation of Graduate Education within the Ministry of Education of Brazil.
M.~K. and Y.~A. acknowledge the support given by the International Relativistic Astrophysics Erasmus Mundus Joint Doctorate Program under the Grants 2013–1471 and 2014–0707, respectively, from EACEA of the European Commission.
M.~M. acknowledges the partial support of the project No. 3101/GF4 IPC-11/2015, and the target program of the Ministry of Education and Science of the Republic of Kazakhstan.

\begin{longtable}{cccccccc}
\caption{List of the the ES of BdHNe considered in this work. It is composed by 161 sources spanning 11 years of Swift/XRT observation activity. In the table we report important observational features: the redshift $z$, our estimates of the LXRE plower-law slope $\alpha$, the late times energy $E_{LT}$, the collimation half-opening angle $\theta$, the isotropic energy $E_{iso}$ of the GRB, the observing instrument in the $\gamma$-ray band, and the correspondent circular (GCN) from which we take the $\gamma$-ray spectral parameters in order to estimate the $E_{iso}$ of the GRB source.\\
$^{(a)}$: in units of $10^{51}$ erg.\\
$^{(b)}$: in units of $10^{52}$ erg.\\
$^{(c)}$: ``Swift'' stays for \textit{Swift}/BAT; ``Fermi'' stays for \textit{Fermi}/GBM; ``KW'' stays for Konus-WIND.\label{EStable}}\\
\hline\hline
$\,\,\,\,\,$ GRB $\,\,\,\,\,$        &      $\,\,\,\,\,\,\,\,\,\,\,$ z $\,\,\,\,\,\,\,\,\,\,\,$         & $\,\,\,\,\,\,\,\,\,\,\,$ $\alpha$ $\,\,\,\,\,\,\,\,\,\,\,$ &    $\,\,\,\,\,\,\,\,\,$  $E_{LT}^{(a)}$ $\,\,\,\,\,\,\,\,\,$   & $\,\,\,\,\,\,\,\,\,$ $\theta$ $\,\,\,$  ($^\circ$) $\,\,\,\,\,\,\,\,\,$ &  $\,\,\,\,\,\,\,\,\,$ $E_{iso}^{(b)} $\,\,\,\,\,\,\,\,\,$ $  &$\,\,\,\,\,$  Instrument$^{(c)}$ $\,\,\,\,\,$ & $\,\,$ GCN $\,\,$  \\
\hline
\endfirsthead
\caption{continued.}\\
\hline\hline
GRB        &       z         & $\alpha$ &      $E_{LT}^{(a)}$   & $\theta$ ($^\circ$) &  $E_{iso}^{(b)} $  & Instrument$^{(c)}$ & GCN  \\
\hline
\endhead
\hline
\endfoot
050315A$^{\,^{\,^{\,^{\,}}}}$	&	1.95	&	0.838 	&	15.6 	&	10.1 	&	8.32 	&	Swift	&	3099	\\
050318A$^{\,^{\,^{\,^{\,}}}}$	&	1.44	&	1.74 	&	0.442 	&	63.3 	&	3.70 	&	Swift	&	3134	\\
050319A$^{\,^{\,^{\,^{\,}}}}$	&	3.24	&	1.27 	&	12.3 	&	11.4 	&	20.2 	&	Swift	&	3119	\\
050401A$^{\,^{\,^{\,^{\,}}}}$	&	2.9	&	1.59 	&	3.82 	&	20.6 	&	92.0 	&	KW	&	3179	\\
050408A$^{\,^{\,^{\,^{\,}}}}$	&	1.24	&	1.14 	&	2.19 	&	27.2 	&	10.8 	&	HETE	&	3188	\\
050505A$^{\,^{\,^{\,^{\,}}}}$	&	4.27	&	1.41 	&	8.66 	&	13.6 	&	44.6 	&	Swift	&	3364	\\
050525A$^{\,^{\,^{\,^{\,}}}}$	&	0.606	&	1.47 	&	0.243 	&	90.0 	&	2.31 	&	KW	&	3474	\\
050730A$^{\,^{\,^{\,^{\,}}}}$	&	3.97	&	2.42 	&	1.74 	&	30.7 	&	42.8 	&	Swift	&	3715	\\
050802A$^{\,^{\,^{\,^{\,}}}}$	&	1.71	&	1.55 	&	1.24 	&	36.5 	&	7.46 	&	Swift	&	3737	\\
050814A$^{\,^{\,^{\,^{\,}}}}$	&	5.3	&	2.23 	&	2.90 	&	23.6 	&	26.8 	&	Swift	&	3803	\\
050820A$^{\,^{\,^{\,^{\,}}}}$	&	2.61	&	1.23 	&	22.1 	&	8.51 	&	39.0 	&	KW	&	3852	\\
050922C$^{\,^{\,^{\,^{\,}}}}$	&	2.2	&	1.57 	&	0.974 	&	41.4 	&	19.8 	&	KW	&	4030	\\
051109A$^{\,^{\,^{\,^{\,}}}}$	&	2.35	&	1.20 	&	9.09 	&	13.3 	&	18.3 	&	KW	&	4238	\\
060108A$^{\,^{\,^{\,^{\,}}}}$	&	2.03	&	0.830 	&	3.40 	&	21.8 	&	1.50 	&	Swift	&	4445	\\
060115A$^{\,^{\,^{\,^{\,}}}}$	&	3.53	&	1.58 	&	2.23 	&	27.0 	&	19.0 	&	Swift	&	4518	\\
060124A$^{\,^{\,^{\,^{\,}}}}$	&	2.296	&	1.30 	&	20.7 	&	8.79 	&	26.0 	&	KW	&	4599	\\
060202A$^{\,^{\,^{\,^{\,}}}}$	&	0.783	&	1.03 	&	4.57 	&	18.8 	&	1.90 	&	Swift	&	4635	\\
060206A$^{\,^{\,^{\,^{\,}}}}$	&	4.05	&	1.44 	&	28.8 	&	7.45 	&	3.90 	&	Swift	&	4697	\\
060210A$^{\,^{\,^{\,^{\,}}}}$	&	3.91	&	1.75 	&	25.9 	&	7.86 	&	120. 	&	Swift	&	4734	\\
060418A$^{\,^{\,^{\,^{\,}}}}$	&	1.49	&	1.54 	&	0.614 	&	52.9 	&	12.0 	&	KW	&	4989	\\
060502A$^{\,^{\,^{\,^{\,}}}}$	&	1.51	&	1.26 	&	2.96 	&	23.4 	&	11.0 	&	Swift	&	5053	\\
060510B$^{\,^{\,^{\,^{\,}}}}$	&	4.9	&	1.54 	&	3.39 	&	21.8 	&	66.0 	&	Swift	&	5107	\\
060512A$^{\,^{\,^{\,^{\,}}}}$	&	2.1	&	1.24 	&	0.356 	&	71.5 	&	2.38 	&	Swift	&	5124	\\
060526A$^{\,^{\,^{\,^{\,}}}}$	&	3.21	&	2.27 	&	1.36 	&	34.7 	&	5.40 	&	Swift	&	5174	\\
060605A$^{\,^{\,^{\,^{\,}}}}$	&	3.8	&	2.00 	&	0.524 	&	57.6 	&	10.1 	&	Swift	&	5231	\\
060607A$^{\,^{\,^{\,^{\,}}}}$	&	3.082	&	3.04 	&	0.481 	&	60.4 	&	34.0 	&	Swift	&	5242	\\
060707A$^{\,^{\,^{\,^{\,}}}}$	&	3.43	&	1.18 	&	11.4 	&	11.8 	&	5.30 	&	Swift	&	5289	\\
060708A$^{\,^{\,^{\,^{\,}}}}$	&	1.92	&	1.21 	&	1.22 	&	36.8 	&	2.20 	&	Swift	&	5295	\\
060714A$^{\,^{\,^{\,^{\,}}}}$	&	2.71	&	1.47 	&	1.93 	&	29.0 	&	19.0 	&	Swift	&	5334	\\
060729A$^{\,^{\,^{\,^{\,}}}}$	&	0.54	&	1.31 	&	3.36 	&	21.9 	&	1.60 	&	Swift	&	5370	\\
060814A$^{\,^{\,^{\,^{\,}}}}$	&	0.84	&	1.16 	&	1.69 	&	31.1 	&	6.30 	&	KW	&	5460	\\
060906A$^{\,^{\,^{\,^{\,}}}}$	&	3.685	&	1.47 	&	0.482 	&	60.3 	&	25.9 	&	Swift	&	5534	\\
061007A$^{\,^{\,^{\,^{\,}}}}$	&	1.261	&	1.68 	&	0.477 	&	60.6 	&	88.0 	&	KW	&	5722	\\
061121A$^{\,^{\,^{\,^{\,}}}}$	&	1.314	&	1.50 	&	3.75 	&	20.7 	&	27.0 	&	KW	&	5837	\\
061126A$^{\,^{\,^{\,^{\,}}}}$	&	1.159	&	1.30 	&	3.06 	&	23.0 	&	8.10 	&	Swift	&	5860	\\
061222A$^{\,^{\,^{\,^{\,}}}}$	&	2.088	&	1.52 	&	18.7 	&	9.25 	&	33.0 	&	KW	&	5984	\\
070110A$^{\,^{\,^{\,^{\,}}}}$	&	2.35	&	1.10 	&	3.20 	&	22.5 	&	11.9 	&	Swift	&	6007	\\
070306A$^{\,^{\,^{\,^{\,}}}}$	&	1.5	&	1.58 	&	8.22 	&	14.0 	&	15.5 	&	Swift	&	6173	\\
070318A$^{\,^{\,^{\,^{\,}}}}$	&	0.84	&	1.42 	&	1.97 	&	28.8 	&	4.37 	&	Swift	&	6212	\\
070508A$^{\,^{\,^{\,^{\,}}}}$	&	0.82	&	1.60 	&	1.08 	&	39.1 	&	7.57 	&	KW	&	6403	\\
070529A$^{\,^{\,^{\,^{\,}}}}$	&	2.5	&	1.22 	&	1.79 	&	30.2 	&	31.9 	&	Swift	&	6468	\\
070802A$^{\,^{\,^{\,^{\,}}}}$	&	2.45	&	1.41 	&	0.633 	&	52.0 	&	1.64 	&	Swift	&	6699	\\
071003A$^{\,^{\,^{\,^{\,}}}}$	&	1.6	&	1.75 	&	1.61 	&	31.9 	&	35.8 	&	KW	&	6849	\\
080210A$^{\,^{\,^{\,^{\,}}}}$	&	2.64	&	1.38 	&	1.26 	&	36.2 	&	11.8 	&	Swift	&	7289	\\
080310A$^{\,^{\,^{\,^{\,}}}}$	&	2.43	&	1.53 	&	1.34 	&	35.1 	&	20.0 	&	Swift	&	7402	\\
080319B$^{\,^{\,^{\,^{\,}}}}$	&	0.94	&	1.59 	&	1.99 	&	28.6 	&	122. 	&	KW	&	7482	\\
080319C$^{\,^{\,^{\,^{\,}}}}$	&	1.95	&	1.72 	&	6.40 	&	15.8 	&	14.6 	&	KW	&	7487	\\
080605A$^{\,^{\,^{\,^{\,}}}}$	&	1.64	&	1.59 	&	1.39 	&	34.5 	&	22.1 	&	KW	&	7854	\\
080607A$^{\,^{\,^{\,^{\,}}}}$	&	3.04	&	1.57 	&	1.77 	&	30.4 	&	187. 	&	KW	&	7862	\\
080721A$^{\,^{\,^{\,^{\,}}}}$	&	2.6	&	1.62 	&	8.93 	&	13.4 	&	132. 	&	KW	&	7995	\\
080804A$^{\,^{\,^{\,^{\,}}}}$	&	2.2	&	1.68 	&	1.08 	&	39.3 	&	56.9 	&	Swift	&	8067	\\
080805A$^{\,^{\,^{\,^{\,}}}}$	&	1.51	&	1.07 	&	1.21 	&	36.9 	&	9.96 	&	Swift	&	8068	\\
080810A$^{\,^{\,^{\,^{\,}}}}$	&	3.35	&	1.57 	&	1.77 	&	30.4 	&	76.1 	&	KW+Swift	&	8101	\\
080905B$^{\,^{\,^{\,^{\,}}}}$	&	2.37	&	1.51 	&	4.67 	&	18.6 	&	9.85 	&	Fermi	&	8205	\\
080916C$^{\,^{\,^{\,^{\,}}}}$	&	4.35	&	1.29 	&	12.1 	&	11.5 	&	242. 	&	Fermi	&	8278	\\
080928A$^{\,^{\,^{\,^{\,}}}}$	&	1.69	&	1.69 	&	0.564 	&	55.3 	&	4.93 	&	Fermi	&	8316	\\
081008A$^{\,^{\,^{\,^{\,}}}}$	&	1.97	&	1.80 	&	0.596 	&	53.7 	&	9.34 	&	Swift	&	8351	\\
081028A$^{\,^{\,^{\,^{\,}}}}$	&	3.04	&	1.66 	&	2.62 	&	24.9 	&	12.0 	&	Swift	&	8428	\\
081109A$^{\,^{\,^{\,^{\,}}}}$	&	0.98	&	1.32 	&	1.07 	&	39.4 	&	1.87 	&	Fermi	&	8505	\\
081121A$^{\,^{\,^{\,^{\,}}}}$	&	2.51	&	1.51 	&	5.77 	&	16.7 	&	24.9 	&	Fermi	&	8546	\\
081203A$^{\,^{\,^{\,^{\,}}}}$	&	2.1	&	2.00 	&	0.446 	&	63.0 	&	35.1 	&	Swift	&	8595	\\
081221A$^{\,^{\,^{\,^{\,}}}}$	&	2.26	&	0.849 	&	24.5 	&	8.09 	&	33.8 	&	Fermi	&	8704	\\
081222A$^{\,^{\,^{\,^{\,}}}}$	&	2.77	&	1.40 	&	4.28 	&	19.4 	&	25.9 	&	Fermi	&	8715	\\
090102A$^{\,^{\,^{\,^{\,}}}}$	&	1.55	&	1.35 	&	1.58 	&	32.2 	&	19.8 	&	KW	&	8776	\\
090313A$^{\,^{\,^{\,^{\,}}}}$	&	3.38	&	2.72 	&	10.4 	&	12.4 	&	13.1 	&	Swift	&	8986	\\
090328A$^{\,^{\,^{\,^{\,}}}}$	&	0.736	&	1.84 	&	1.42 	&	34.0 	&	19.5 	&	Fermi	&	9056	\\
090418A$^{\,^{\,^{\,^{\,}}}}$	&	1.61	&	1.68 	&	0.967 	&	41.5 	&	15.0 	&	KW	&	9171	\\
090423A$^{\,^{\,^{\,^{\,}}}}$	&	8.2	&	1.57 	&	1.29 	&	35.7 	&	10.9 	&	Fermi	&	9229	\\
090424A$^{\,^{\,^{\,^{\,}}}}$	&	0.544	&	1.18 	&	1.76 	&	30.5 	&	4.51 	&	Fermi	&	9230	\\
090516A$^{\,^{\,^{\,^{\,}}}}$	&	3.9	&	1.40 	&	4.52 	&	18.9 	&	65.0 	&	Fermi	&	9413	\\
090618A$^{\,^{\,^{\,^{\,}}}}$	&	0.54	&	1.49 	&	1.44 	&	33.8 	&	25.4 	&	Fermi	&	9535	\\
090715B$^{\,^{\,^{\,^{\,}}}}$	&	3.	&	1.18 	&	4.08 	&	19.9 	&	22.1 	&	KW	&	9679	\\
090809A$^{\,^{\,^{\,^{\,}}}}$	&	2.74	&	1.57 	&	0.727 	&	48.3 	&	4.98 	&	Swift	&	9756	\\
090812A$^{\,^{\,^{\,^{\,}}}}$	&	2.45	&	1.23 	&	5.39 	&	17.3 	&	41.7 	&	KW+Swift	&	9821	\\
090902B$^{\,^{\,^{\,^{\,}}}}$	&	1.82	&	1.22 	&	6.15 	&	16.2 	&	403. 	&	Fermi	&	9866	\\
090926A$^{\,^{\,^{\,^{\,}}}}$	&	2.11	&	1.50 	&	5.29 	&	17.4 	&	217. 	&	Fermi	&	9933	\\
091003A$^{\,^{\,^{\,^{\,}}}}$	&	0.9	&	1.36 	&	1.32 	&	35.3 	&	10.9 	&	Fermi	&	9983	\\
091020A$^{\,^{\,^{\,^{\,}}}}$	&	1.71	&	1.30 	&	2.19 	&	27.3 	&	7.63 	&	Fermi	&	10095	\\
091029A$^{\,^{\,^{\,^{\,}}}}$	&	2.75	&	1.60 	&	5.81 	&	16.6 	&	7.52 	&	Swift	&	10103	\\
091127A$^{\,^{\,^{\,^{\,}}}}$	&	0.49	&	1.32 	&	1.36 	&	34.9 	&	1.79 	&	Fermi	&	10204	\\
091208B$^{\,^{\,^{\,^{\,}}}}$	&	1.06	&	1.09 	&	1.15 	&	38.0 	&	2.03 	&	Fermi	&	10266	\\
100302A$^{\,^{\,^{\,^{\,}}}}$	&	4.81	&	0.812 	&	5.67 	&	16.8 	&	5.24 	&	Swift	&	10462	\\
100425A$^{\,^{\,^{\,^{\,}}}}$	&	1.75	&	1.20 	&	0.767 	&	46.9 	&	2.84 	&	Swift	&	10685	\\
100513A$^{\,^{\,^{\,^{\,}}}}$	&	4.77	&	1.44 	&	3.74 	&	20.8 	&	25.7 	&	Swift	&	10753	\\
100621A$^{\,^{\,^{\,^{\,}}}}$	&	0.542	&	1.12 	&	8.52 	&	13.7 	&	6.93 	&	KW	&	10882	\\
100814A$^{\,^{\,^{\,^{\,}}}}$	&	1.44	&	1.64 	&	14.3 	&	10.6 	&	15.8 	&	Fermi	&	11099	\\
100901A$^{\,^{\,^{\,^{\,}}}}$	&	1.41	&	1.36 	&	7.89 	&	14.3 	&	8.09 	&	Swift	&	11169	\\
100906A$^{\,^{\,^{\,^{\,}}}}$	&	1.73	&	1.87 	&	1.32 	&	35.3 	&	22.5 	&	Fermi	&	11248	\\
110128A$^{\,^{\,^{\,^{\,}}}}$	&	2.339	&	1.16 	&	0.620 	&	52.6 	&	3.60 	&	Fermi	&	11628	\\
110205A$^{\,^{\,^{\,^{\,}}}}$	&	2.22	&	1.59 	&	0.913 	&	42.8 	&	41.0 	&	KW	&	11659	\\
110213A$^{\,^{\,^{\,^{\,}}}}$	&	1.46	&	1.81 	&	2.33 	&	26.4 	&	7.00 	&	Fermi	&	11727	\\
110422A$^{\,^{\,^{\,^{\,}}}}$	&	1.77	&	1.24 	&	6.03 	&	16.3 	&	62.0 	&	KW	&	11971	\\
110503A$^{\,^{\,^{\,^{\,}}}}$	&	1.613	&	1.36 	&	2.45 	&	25.7 	&	19.0 	&	KW	&	12008	\\
110715A$^{\,^{\,^{\,^{\,}}}}$	&	0.82	&	1.69 	&	7.33 	&	14.8 	&	5.60 	&	KW	&	12166	\\
110731A$^{\,^{\,^{\,^{\,}}}}$	&	2.83	&	1.22 	&	6.31 	&	16.0 	&	42.0 	&	Fermi	&	12221	\\
110808A$^{\,^{\,^{\,^{\,}}}}$	&	1.348	&	1.33 	&	0.588 	&	54.1 	&	6.10 	&	Swift	&	12262	\\
110918A$^{\,^{\,^{\,^{\,}}}}$	&	0.982	&	1.64 	&	19.5 	&	9.07 	&	190. 	&	KW	&	12362	\\
111008A$^{\,^{\,^{\,^{\,}}}}$	&	4.9898	&	1.66 	&	8.76 	&	13.5 	&	82.0 	&	KW	&	12433	\\
111123A$^{\,^{\,^{\,^{\,}}}}$	&	3.1516	&	1.55 	&	2.82 	&	24.0 	&	70.0 	&	Swift	&	12598	\\
111209A$^{\,^{\,^{\,^{\,}}}}$	&	0.67	&	1.54 	&	1.45 	&	33.6 	&	67.0 	&	KW	&	12663	\\
111228A$^{\,^{\,^{\,^{\,}}}}$	&	0.716	&	1.23 	&	1.24 	&	36.4 	&	4.10 	&	Fermi	&	12744	\\
120119A$^{\,^{\,^{\,^{\,}}}}$	&	1.73	&	1.28 	&	1.68 	&	31.2 	&	37.5 	&	Fermi	&	12874	\\
120326A$^{\,^{\,^{\,^{\,}}}}$	&	1.8	&	1.87 	&	9.97 	&	12.7 	&	3.70 	&	Fermi	&	13145	\\
120327A$^{\,^{\,^{\,^{\,}}}}$	&	2.81	&	1.58 	&	1.33 	&	35.1 	&	36.5 	&	Swift	&	13137	\\
120711A$^{\,^{\,^{\,^{\,}}}}$	&	1.41	&	1.58 	&	18.1 	&	9.40 	&	166. 	&	Fermi	&	13437	\\
120712A$^{\,^{\,^{\,^{\,}}}}$	&	4.17	&	1.37 	&	2.90 	&	23.7 	&	18.3 	&	Fermi	&	13469	\\
120909A$^{\,^{\,^{\,^{\,}}}}$	&	3.93	&	1.30 	&	11.8 	&	11.7 	&	73.2 	&	Fermi	&	13737	\\
120922A$^{\,^{\,^{\,^{\,}}}}$	&	3.1	&	1.19 	&	5.41 	&	17.3 	&	32.0 	&	Fermi	&	13809	\\
121024A$^{\,^{\,^{\,^{\,}}}}$	&	2.3	&	1.44 	&	1.18 	&	37.4 	&	10.7 	&	Swift	&	13899	\\
121027A$^{\,^{\,^{\,^{\,}}}}$	&	1.77	&	1.28 	&	15.8 	&	10.1 	&	6.55 	&	Swift	&	13910	\\
121128A$^{\,^{\,^{\,^{\,}}}}$	&	2.2	&	1.48 	&	1.85 	&	29.7 	&	13.2 	&	Fermi	&	14012	\\
121217A$^{\,^{\,^{\,^{\,}}}}$	&	3	&	1.26 	&	30.6 	&	7.23 	&	24.2 	&	Fermi	&	14094	\\
130418A$^{\,^{\,^{\,^{\,}}}}$	&	1.218	&	1.52 	&	0.265 	&	85.4 	&	9.90 	&	KW	&	14417	\\
130420A$^{\,^{\,^{\,^{\,}}}}$	&	1.297	&	1.25 	&	1.32 	&	35.4 	&	7.74 	&	Fermi	&	14429	\\
130427A$^{\,^{\,^{\,^{\,}}}}$	&	0.338	&	1.26 	&	4.17 	&	19.7 	&	105. 	&	Fermi	&	14473	\\
130427B$^{\,^{\,^{\,^{\,}}}}$	&	2.78	&	1.85 	&	0.275 	&	83.4 	&	5.04 	&	Swift	&	14469	\\
130505A$^{\,^{\,^{\,^{\,}}}}$	&	2.27	&	1.50 	&	16.0 	&	10.0 	&	347. 	&	KW	&	14575	\\
130514A$^{\,^{\,^{\,^{\,}}}}$	&	3.6	&	1.56 	&	5.16 	&	17.7 	&	52.4 	&	KW+Swift	&	14702	\\
130528A$^{\,^{\,^{\,^{\,}}}}$	&	1.25	&	1.02 	&	2.03 	&	28.3 	&	4.40 	&	Fermi	&	14729	\\
130606A$^{\,^{\,^{\,^{\,}}}}$	&	5.91	&	1.91 	&	1.24 	&	36.4 	&	28.3 	&	KW	&	14808	\\
130610A$^{\,^{\,^{\,^{\,}}}}$	&	2.092	&	1.47 	&	0.472 	&	61.0 	&	6.99 	&	Fermi	&	14858	\\
130701A$^{\,^{\,^{\,^{\,}}}}$	&	1.155	&	1.20 	&	0.639 	&	51.7 	&	2.60 	&	KW	&	14958	\\
130907A$^{\,^{\,^{\,^{\,}}}}$	&	1.238	&	1.68 	&	7.02 	&	15.1 	&	304. 	&	KW	&	15203	\\
130925A$^{\,^{\,^{\,^{\,}}}}$	&	0.347	&	1.32 	&	2.85 	&	23.8 	&	18.4 	&	Fermi	&	15255	\\
131030A$^{\,^{\,^{\,^{\,}}}}$	&	1.293	&	1.27 	&	2.97 	&	23.4 	&	173. 	&	KW	&	15413	\\
131105A$^{\,^{\,^{\,^{\,}}}}$	&	1.686	&	1.24 	&	1.75 	&	30.6 	&	34.7 	&	Fermi	&	15455	\\
131108A$^{\,^{\,^{\,^{\,}}}}$	&	2.4	&	1.72 	&	0.932 	&	42.3 	&	73.0 	&	Fermi	&	15477	\\
131117A$^{\,^{\,^{\,^{\,}}}}$	&	4.042	&	1.31 	&	0.816 	&	45.4 	&	1.02 	&	Swift	&	15499	\\
131231A$^{\,^{\,^{\,^{\,}}}}$	&	0.642	&	1.44 	&	3.39 	&	21.8 	&	22.2 	&	Fermi	&	15644	\\
140206A$^{\,^{\,^{\,^{\,}}}}$	&	2.74	&	1.43 	&	13.5 	&	10.9 	&	35.9 	&	Fermi	&	15796	\\
140213A$^{\,^{\,^{\,^{\,}}}}$	&	1.2076	&	2.21 	&	16.2 	&	9.94 	&	9.93 	&	Fermi	&	15833	\\
140226A$^{\,^{\,^{\,^{\,}}}}$	&	1.98	&	1.58 	&	0.762 	&	47.1 	&	5.80 	&	KW	&	15889	\\
140304A$^{\,^{\,^{\,^{\,}}}}$	&	5.283	&	1.44 	&	7.14 	&	15.0 	&	13.7 	&	Fermi	&	15923	\\
140311A$^{\,^{\,^{\,^{\,}}}}$	&	4.954	&	1.49 	&	1.18 	&	37.5 	&	11.6 	&	Swift	&	15962	\\
140419A$^{\,^{\,^{\,^{\,}}}}$	&	3.956	&	1.84 	&	9.32 	&	13.1 	&	186. 	&	KW	&	16134	\\
140423A$^{\,^{\,^{\,^{\,}}}}$	&	3.26	&	1.31 	&	4.11 	&	19.8 	&	65.3 	&	Fermi	&	16152	\\
140506A$^{\,^{\,^{\,^{\,}}}}$	&	0.889	&	0.924 	&	2.78 	&	24.2 	&	1.12 	&	Fermi	&	16220	\\
140508A$^{\,^{\,^{\,^{\,}}}}$	&	1.027	&	1.40 	&	5.46 	&	17.2 	&	23.9 	&	Fermi	&	16224	\\
140509A$^{\,^{\,^{\,^{\,}}}}$	&	2.4	&	1.46 	&	0.402 	&	66.7 	&	9.14 	&	Swift	&	16240	\\
140512A$^{\,^{\,^{\,^{\,}}}}$	&	0.725	&	1.62 	&	2.22 	&	27.1 	&	7.76 	&	Fermi	&	16262	\\
140614A$^{\,^{\,^{\,^{\,}}}}$	&	4.233	&	1.24 	&	2.00 	&	28.5 	&	7.30 	&	Swift	&	16402	\\
140620A$^{\,^{\,^{\,^{\,}}}}$	&	2.04	&	1.43 	&	5.16 	&	17.7 	&	6.22 	&	Fermi	&	16426	\\
140629A$^{\,^{\,^{\,^{\,}}}}$	&	2.275	&	1.48 	&	0.868 	&	44.0 	&	6.15 	&	KW	&	16495	\\
140703A$^{\,^{\,^{\,^{\,}}}}$	&	3.14	&	2.21 	&	1.82 	&	30.0 	&	1.72 	&	Fermi	&	16512	\\
140907A$^{\,^{\,^{\,^{\,}}}}$	&	1.21	&	0.888 	&	2.68 	&	24.6 	&	2.29 	&	Fermi	&	16798	\\
141026A$^{\,^{\,^{\,^{\,}}}}$	&	3.35	&	2.24 	&	34.8 	&	6.78 	&	7.17 	&	Swift	&	16960	\\
141109A$^{\,^{\,^{\,^{\,}}}}$	&	2.993	&	1.31 	&	7.85 	&	14.3 	&	5.05 	&	KW	&	17055	\\
141121A$^{\,^{\,^{\,^{\,}}}}$	&	1.47	&	1.67 	&	6.56 	&	15.6 	&	14.6 	&	KW	&	17108	\\
141221A$^{\,^{\,^{\,^{\,}}}}$	&	1.452	&	1.18 	&	0.970 	&	41.5 	&	1.94 	&	Fermi	&	17216	\\
150206A$^{\,^{\,^{\,^{\,}}}}$	&	2.087	&	1.28 	&	10.1 	&	12.6 	&	151. 	&	KW	&	17427	\\
150314A$^{\,^{\,^{\,^{\,}}}}$	&	1.758	&	1.57 	&	3.08 	&	22.9 	&	98.1 	&	Fermi	&	17579	\\
150403A$^{\,^{\,^{\,^{\,}}}}$	&	2.06	&	1.61 	&	26.7 	&	7.74 	&	91.0 	&	Fermi	&	17674	\\
150514A$^{\,^{\,^{\,^{\,}}}}$	&	0.807	&	1.35 	&	0.378 	&	69.1 	&	1.14 	&	Fermi	&	17819	\\
150821A$^{\,^{\,^{\,^{\,}}}}$	&	0.755	&	1.20 	&	0.884 	&	43.5 	&	15.1 	&	Fermi	&	18190	\\
150910A$^{\,^{\,^{\,^{\,}}}}$	&	1.36	&	1.50 	&	0.415 	&	65.6 	&	22.3 	&	Swift	&	18268	\\
151021A$^{\,^{\,^{\,^{\,}}}}$	&	2.33	&	1.47 	&	2.69 	&	24.6 	&	211. 	&	KW	&	18433	\\
151027A$^{\,^{\,^{\,^{\,}}}}$	&	0.81	&	1.69 	&	2.04 	&	28.3 	&	3.86 	&	Fermi	&	18492	\\
151027B$^{\,^{\,^{\,^{\,}}}}$	&	4.063	&	1.75 	&	4.19 	&	19.6 	&	19.1 	&	Swift	&	18514	\\
151111A$^{\,^{\,^{\,^{\,}}}}$	&	3.5	&	1.29 	&	0.871 	&	43.9 	&	5.13 	&	Fermi	&	18582	\\
151112A$^{\,^{\,^{\,^{\,}}}}$	&	4.1	&	1.54 	&	3.31 	&	22.1 	&	12.4 	&	Swift	&	18593	\\
151215A$^{\,^{\,^{\,^{\,}}}}$	&	2.59	&	1.07 	&	1.34 	&	35.1 	&	1.97 	&	Swift	&	18699	\\
\end{longtable}

\end{document}